\documentclass[superscriptaddress,amsmath,amssymb,aps,prb,twocolumn,nofootinbib,nolongbibliography]{revtex4-2}
\usepackage{natbib}
\usepackage{graphicx}
\usepackage{graphics}
\usepackage{dcolumn}          
\usepackage{bm}               
\usepackage{upgreek}
\usepackage{booktabs} 				
\usepackage{tabularx}
\usepackage{array}
\usepackage[colorlinks=true,urlcolor=blue,breaklinks=true]{hyperref}
\usepackage[justification=justified,singlelinecheck=false]{caption}
\usepackage{afterpage}
\usepackage{xcolor}

\usepackage{orcidlink}

\usepackage{xr}

\usepackage{siunitx}
\usepackage{stackengine}
\usepackage{svg}
\usepackage{graphicx,subfigure}
\usepackage{placeins}
\usepackage{bibunits}

\usepackage{ragged2e} 

\usepackage{titlesec}
\begin{document}
\title{Interplay between Aharonov--Bohm and Altshuler--Aronov--Spivak oscillations in phase-pure GaAs/InAs core/shell nanowires of different lengths}

\author{Farah Basari\'{c}\,\orcidlink{0009-0000-8721-3129}}
\thanks{These authors contributed equally to this work}
\affiliation{Peter Gr\"unberg Institut (PGI-9), Forschungszentrum J\"ulich, 52425 J\"ulich, Germany}
\affiliation{JARA-Fundamentals of Future Information Technology, J\"ulich-Aachen Research Alliance, Forschungszentrum J\"ulich and RWTH Aachen University, 52425 J\"ulich, Germany}
\author{Kaiwen Wang}
\thanks{These authors contributed equally to this work}
\affiliation{Peter Gr\"unberg Institut (PGI-9), Forschungszentrum J\"ulich, 52425 J\"ulich, Germany}
\affiliation{JARA-Fundamentals of Future Information Technology, J\"ulich-Aachen Research Alliance, Forschungszentrum J\"ulich and RWTH Aachen University, 52425 J\"ulich, Germany}
\author{Tudor-Gabriel Dumitru\,\orcidlink{0009-0001-1970-0766}}
\affiliation{Department of Engineering, Reykjavik University, Menntavegur 1, IS-102 Reykjavik, Iceland}
\author{Andrei Manolescu\,\orcidlink{0000-0002-0713-4664}}
\affiliation{Department of Engineering, Reykjavik University, Menntavegur 1, IS-102 Reykjavik, Iceland}
\author{Francisco Alvarado Cesar\,\orcidlink{0009-0003-6220-5642}}
\thanks{Present address: Facility for Electron Microscopy, The University of Birmingham, Edgbaston, Birmingham, B15 2TT, United Kingdom}
\affiliation{Department of Physics, University of Warwick, Coventry CV4 7AL, UK}
\author{Ana M. Sanchez} \affiliation{Department of Physics, University of Warwick, Coventry CV4 7AL, UK}
\author{Christoph Krause}
\affiliation{Peter Gr\"unberg Institut (PGI-10), Forschungszentrum J\"ulich, 52425 J\"ulich, Germany}
\author{Detlev Gr\"utzmacher\,\orcidlink{0000-0001-6290-9672}}
\affiliation{Peter Gr\"unberg Institut (PGI-9), Forschungszentrum J\"ulich, 52425 J\"ulich, Germany}
\affiliation{Peter Gr\"unberg Institut (PGI-10), Forschungszentrum J\"ulich, 52425 J\"ulich, Germany}
\affiliation{JARA-Fundamentals of Future Information Technology, J\"ulich-Aachen Research Alliance, Forschungszentrum J\"ulich and RWTH Aachen University, 52425 J\"ulich, Germany}
\author{Alexander Pawlis\,\orcidlink{0000-0002-3394-0707}}
\affiliation{Peter Gr\"unberg Institut (PGI-9), Forschungszentrum J\"ulich, 52425 J\"ulich, Germany}
\affiliation{Peter Gr\"unberg Institut (PGI-10), Forschungszentrum J\"ulich, 52425 J\"ulich, Germany}
\affiliation{JARA-Fundamentals of Future Information Technology, J\"ulich-Aachen Research Alliance, Forschungszentrum J\"ulich and RWTH Aachen University, 52425 J\"ulich, Germany}
\author{Thomas Sch\"apers\,\orcidlink{0000-0001-7861-5003}}
\email{th.schaepers@fz-juelich.de}
\affiliation{Peter Gr\"unberg Institut (PGI-9), Forschungszentrum J\"ulich, 52425 J\"ulich, Germany}
\affiliation{JARA-Fundamentals of Future Information Technology, J\"ulich-Aachen Research Alliance, Forschungszentrum J\"ulich and RWTH Aachen University, 52425 J\"ulich, Germany}
\affiliation{Department of Material Science, Tohoku University, Aobayama 2, Aoba-ku, Sendai 980-8579, Japan}
\hyphenation{}
\date{\today}

\begin{abstract}
In GaAs/InAs core/shell nanowires, comprising a tubular conducting shell, interference phenomena observed under an axial field and originating from closed-loop states encircling the insulating core, provide an ideal platform for superconducting quantum devices that utilize effects such as Aharonov--Bohm or Altshuler--Aronov--Spivak-type conductance oscillations. Both effects are different in nature with respect to phase rigidity because of interference of non-time-reversed or time-reversed paths, respectively. Since their occurrence is largely governed by averaging effects, which depend on sample dimensions and the transport regime, we present a systematic study of flux-periodic oscillations of phase-pure zinc-blende GaAs/InAs core/shell nanowires as a function of gate voltage for samples with different contact separation lengths. Our analysis shows that with increasing contact separation length, averaging effects result in gradual reduction of $h/e$-periodic Aharonov--Bohm-type oscillations, while the $h/2e$-periodic Altshuler--Aronov--Spivak oscillations and its $h/4e$-periodic higher harmonics are enhanced. The additional phase rigidity seen in the $h/3e$-periodic oscillations is attributed to phase rigidity propagating from the neighbouring lower harmonics. Our tight-binding transport simulations on nanowires of different lengths which contain only a few scattering centers confirm the experimental observations regarding the different harmonics and their phase rigidity. Together, our experimental and simulation findings indicate quasi-ballistic transport with persistent Aharonov--Bohm-, and phase-rigid Altshuler--Aronov--Spivak-type oscillations despite few scattering centers.
\end{abstract}
\maketitle

\section{Introduction}
\begin{bibunit}[]
Semiconductor nanowires with a tubular conducting shell provide a particularly suitable platform for superconducting quantum devices that exploit electron interference phenomena \cite{Guel2014a,Gharavi2015,Haas18,Stampfer2022,Zellekens2025}. In this work, we investigate GaAs/InAs core/shell nanowires \cite{Bloemers2013,Guel2014,Haas2016,Haas2017,Basaric2025}, which exhibit a rich mesoscopic behavior, while simultaneously offering a highly tunable system for investigating coherent electron dynamics and device functionality with strong potential for future application in more complex devices for classical and quantum computing \cite{deLange2015,Larsen2015,Tosi2019,Metzger2021,Zellekens2022,Yazdani2023,Sarma2015}. A notable example is the realization of a core/shell nanowire-based Josephson junction \cite{Zellekens2025}, where strong coupling between superconducting electrodes and the semiconducting nanowire weak link is essential. GaAs/InAs core/shell nanowires are promising candidates for such applications because they enable the precise control of the supercurrent in the Josephson junction via field-effect. Additionally, the band offset at the interface between the high-band-gap core and the low-band-gap shell effectively confines charge carriers in a tubular conductor  \cite{Tserkovnyak2006,Rosdahl2014,Ferrari2008,Manolescu2016}. As a result, the electron wavefunction is localized near the interface with the superconducting electrodes. Studying interference effects in this respect offers direct insight into the phase coherence and transport properties of charge carriers in tubular conducting channels, as it allows for further development of quantum information devices based on such nanostructures.

When subjected to an axial magnetic field, GaAs/InAs core/shell nanowires exhibit such interference effects, manifested as magnetic flux quantum $h/e$-periodic Aharonov--Bohm- (AB) \cite{Aharonov1959} and $h/2e$-periodic Altshuler–Aronov–Spivak-type (AAS) conductance oscillations \cite{Altshuler1981}, where $h$ is the Planck constant, and $e$ the elementary charge. In case of planar ring structures based on semiconducting \cite{Mailly1993,Datta1985}, metallic \cite{Sharvin1981,Aronov1987, CPUmbach,Levy1990,Chandrasekhar1991}, and graphene \cite{Bachtold1999,DauberStampfer17} nanostructures, these effects originate from electron path interference. In contrast to that, the electronic transport in GaAs/InAs nanowires is defined by closed-loop states that encircle the insulating core, also known as angular momentum states \cite{Guel2014,Bloemers2013}. Applying an axial magnetic field, the flux-dependent energy spectrum of their angular momentum states is revealed through $h/e$-periodic conductance oscillations, therefore classified as an AB-type oscillations. Commonly, they have been observed as a dominant effect in GaAs/InAs nanowires. Furthermore, in terms of electron interference framework, AAS oscillations are described as the interference of time-reversed paths. In case of GaAs/InAs nanowires, so far they have been associated with elastic disorder scattering in the diffusive transport regime \cite{Guel2014}. Recently, however, those were also reported in highly-ordered quasiballistic GaAs/InAs nanowires, mediated by a few scattering centers only \cite{Basaric2025}.

Previous interference studies of such phenomena on III–V semiconductor \cite{Bloemers2013,Guel2014,Haas2016,Haas2017,Zellekens2020,Jung2008,Basaric2025}, and topological insulator nanowires \cite{Peng2010,Cho2015,Arango2016,Rosenbach2020} have demonstrated the critical role of system size, crystal structure, and characteristic transport length scales in shaping these interference patterns, as well as determining the crossover conditions between AB and AAS oscillations \cite{Kim2020}. Additional to $h/e$- and $h/2e$-periodic oscillations, experiments on planar quasiballistic ring nanostructures \cite{Hansen2001} and topological insulator nanowires \cite{Dufouleur2013} showcased higher harmonic oscillations, i.e. $h/3e$-, $h/4e$- etc. periodicities. These are enabled due to large phase coherence length, which allows the electron wavefunction to wind around the ring perimeter several times before interference occurs. However, higher harmonics have not been reported yet in tubular semiconductor nanowires.

Recently, we investigated magnetoconductance oscillations in phase-pure GaAs/InAs core/shell nanowires \cite{Basaric2025}. Despite the short nanowire length and reported quasiballistic transport regime, we observed AAS corrections at small magnetic fields. To better understand the relationship between AB- and AAS-type oscillations, the present study investigates the magnetoconductance oscillations in similarly grown phase-pure GaAs/InAs core/shell nanowires with various contact separation lengths ranging from 300\,nm to 2\,{\textmu}m. This allows us to access transport scenarios with different averaging of interference phenomena. We investigate these phenomena via transversal mode of transport by measuring magnetoconductance oscillations under an axial field at low temperatures, and varying the gate voltage which probes the phase of the observed oscillations. We expect AB-type oscillations to progressively average out with increasing contact separation length, while AAS-type oscillations stabilize. Furthermore, we report the observation of higher harmonics of magnetoconductance oscillations and present a study of their phase rigidity and the relevant magnetic field ranges within which they occur. Our experimental results are compared to tight-binding quantum transport simulations of core/shell nanowires with  different lengths. These contain only a few scattering centers based on Landauer--B\"uttiker formalism using KWANT \cite{Groth2014}. The presented simulation results reveal the underlying mechanism behind the observed higher harmonics.

\section{Experimental details}

\begin{figure*}[!ht]
    \centering 
    \includegraphics[width=\textwidth]{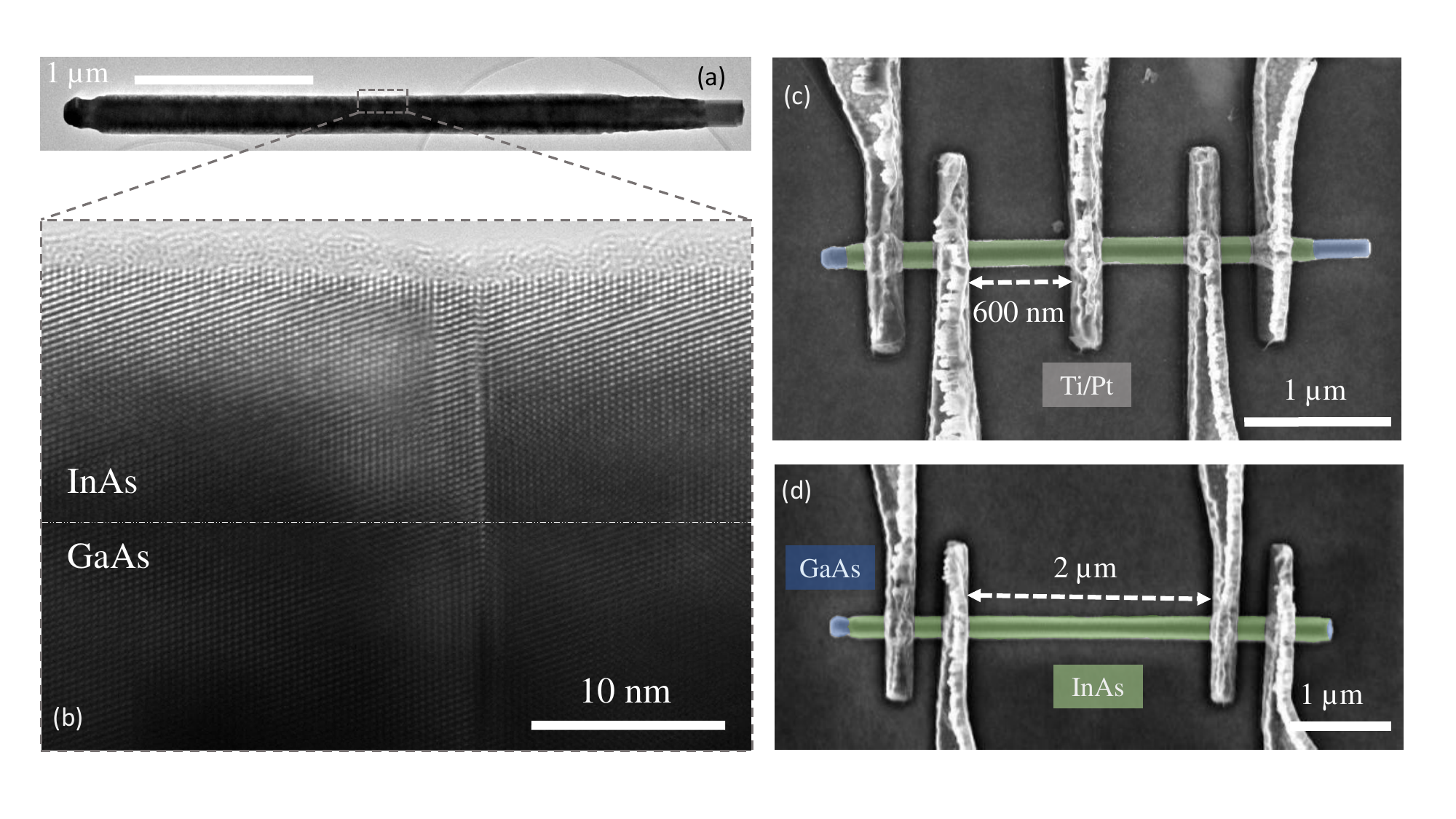}  
    \caption{\justifying(a) Low-magnification TEM image of a representative example of the GaAs/InAs nanowire, featuring (b) high-resolution TEM image of the nanowire middle region. (c),(d) False color SEM image of sample C and E, respectively, emphasizing differences in inner contact separation lengths $L_{\mathrm{c}}$, 600\,nm and 2\,$\mu$m, respectively.}
    \captionsetup{position=bottom}
    \label{fig:Figure-SEM}
\end{figure*}

The GaAs/InAs core/shell nanowires investigated were grown by molecular beam epitaxy (MBE) via the self-catalyzed vapor-liquid-solid technique on pre-structured Si(111) substrates, following the procedure reported by Jansen et al. \cite{Jansen2020}. The growth of the GaAs core resulted in average radius of $65\,$nm, while the average thickness of the InAs shell was 30\,nm. The nanowires had a typical length of 3.8\,{\textmu}m. Details on the growth procedure can be found in the Supplemental Material \cite{Basaric-suppl2026}. Table \ref{tab:Geometry} presents the resulting nanowire geometries of the measured samples.
\begin{table*}[!htb]
    \centering
    \begin{tabular}{c|c|c|c|c|c|c|c|c|c}
        \hline
 Sample & $L_{\mathrm{c}}\,(\mathrm{{\textmu} m})$ & $r_{\mathrm{core}}\,(\mathrm{nm})$ & $t_{\mathrm{shell}}\,$(nm) & $r_{\mathrm{tot}}\,$(nm) & $f_{\mathrm{h/e}}\,(\mathrm{T}^{-1})$  & $f_{\mathrm{h/2e}}\,(\mathrm{T}^{-1})$ & $f_{\mathrm{h/3e}}\,(\mathrm{T}^{-1})$ & $f_{\mathrm{h/4e}}\,(\mathrm{T}^{-1})$ & $r_{\mathrm{h/e}}\,$(nm)\\
\hline
 A & 0.3 & 56 & 37 & 93 & 3.6 & 7.2 & - & - & 75\\ 
 B & 0.5 & 71 & 25 & 96 & 3.3  & 6.7 & - & - & 72\\ 
 C & 0.6 & 65 & 31 & 96 & 3.0 & 5.9 & 8.5 & 12.1 & 69\\ 
 D & 1.8 & 70 & 27 & 97 & 3.0 & 6.0 & 9.0 & 12.0 & 69\\ 
 E & 2.0 & 63 & 27 & 90 & 3.1  & 5.8 & 9.0 & 12.0 & 70\\ 
 \hline
    \end{tabular}
    \caption{\justifying Sample-specific geometric dimensions and corresponding oscillation periodicities, where $L_{c}$ denotes the contact separation of the voltage probes, $r_\mathrm{core}$ the GaAs core radius, $t_\mathrm{shell}$ the InAs shell thickness, $r_\mathrm{tot}$ the total nanowire radius, $f_{h/e}$, $f_{h/2e}$, $f_{h/3e}$, and $f_{h/4e}$ the positions of the $h/e$-, $h/2e$-, $h/3e$-, and $h/4e$-periodic peaks on the frequency scale of the fast Fourier transform analysis (third and fourth harmonics are observed only for samples C, D, and E), and $r_{h/e}$ the corresponding measured wave function localization radius, obtained at the base temperature of 1.3\,K.}
    \label{tab:Geometry}
\end{table*}
A transmission electron microscopy (TEM) analysis was carried out to gain information on the crystal structure of the nanowire. Due to the interplay between hole diameter and droplet size at growth start, the growth of the nanowires was stabilized in the zincblende regime, featuring an excellent degree of phase-purity. Figure~\ref{fig:Figure-SEM} (a) shows a low magnification bright field transmission electron microscopy image (BF-TEM) of a typical nanowire, where a high-resolution TEM image in Fig.~\ref{fig:Figure-SEM} (b) along the $<$110$>$ direction confirms the preferred zincblende crystal structure.

The device fabrication required primarily the precise nanowire transferring technique using a scanning electron microscope-based micromanipulator onto a pre-patterned $n$-doped Si(100) global backgate substrate covered with 150\,nm of $\mathrm{SiO_{2}}$ used as gate dielectric. Next, the contacts were defined using electron beam lithography, followed by Ar$^\mathrm{+}$ cleaning and contact finger metallization with Ti/Au. Representative examples of the measured samples are shown in Figs.~\ref{fig:Figure-SEM} (c) and (d). The low-temperature measurements were carried out in a variable temperature insert cryostat, at the base temperature of 1.3\,K. A four-terminal measurement setup was utilized, except for the shortest sample, which as measured in a three-terminal configuration, by means of applying 20\,nA of current between the two outer contacts, whereas the voltage drop was measured between the two inner ones. The typical contact resistances was in order of 1\,k$\mathrm{\Omega}$. We present results of gate-dependent magnetoconductance measurements on samples with different inner contact separation lengths $L_{\mathrm{c}}$ given in Table \ref{tab:Geometry}. The measurements were carried out under an axial magnetic field, and device gating was realized by a global back-gate.

\section{Results and Discussion}

\begin{figure*}[!htbp]
    \centering 
   \includegraphics[width=\textwidth]{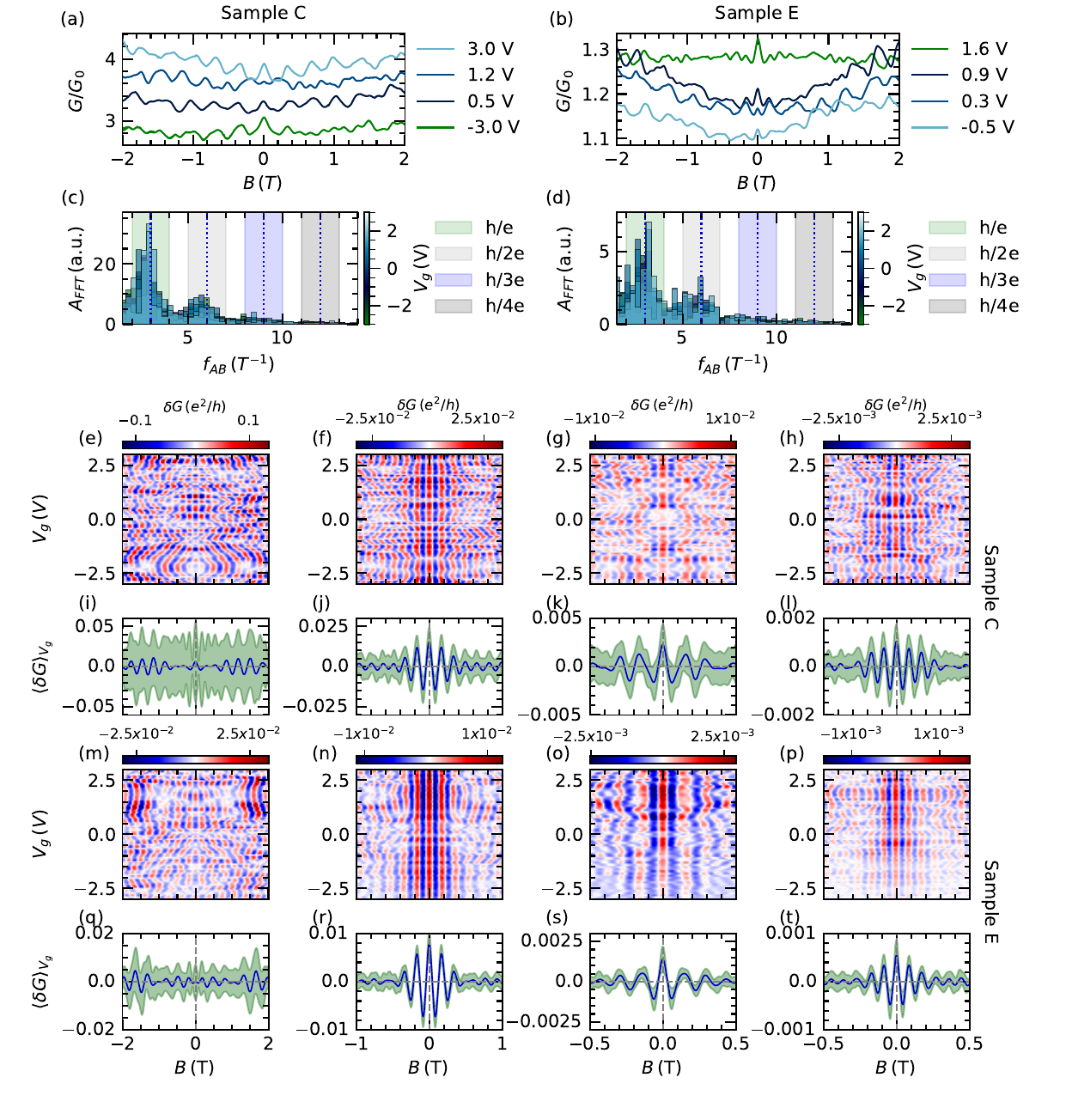}  
    \caption{\justifying Normalized magnetoconductance $G/G_{0}$, with $G_{0}=e^{2}/h$, for various gate voltage $V_{g}$ values, for (a) sample C, and (b) E, respectively. Corresponding fast-Fourier transform of the data shown in (a) and (b), for samples (c) C, and (d) E. Ranges of present harmonics ($h/e$, $h/2e$, $h/3e$, $h/4e$) are presented in color-coded scheme, with mean frequency value marked by dotted line. Colormaps for oscillation amplitude $\delta G$ as a function of gate voltage and magnetic field for sample C, presented for (e) $h/e$-, (f) $h/2e$-, (g) $h/3e$-, and (h) $h/4e$-periodic oscillations defined by the frequency windows marked in (b). Averaged oscillation amplitude $\langle \delta G \rangle_{V_\mathrm{g}}$ over the complete gate voltage range, as a function of magnetic field, presented with bold line, and corresponding standard deviation marked with shaded area for (i) $h/e$-, (j) $h/2e$-, (k) $h/3e$-, and (l) $h/4e$-periodic oscillations. Colormaps in (m), (n), (o) and (p) represent analogous data analysis for sample E, shown for marked frequency ranges in (d). Averaged oscillation amplitude over gate voltage range with standard deviation of different harmonics is similarly shown for sample E in (q), (r), (s) and (t).}
    \captionsetup{position=bottom}
    \label{fig:Figure-Gate}
\end{figure*}

We examined magnetotransport in samples with varying inner contact separation $L_{\mathrm{c}}$, with a focus on samples C and E, which contain short $600\,$nm and long $2\,\mu$m contact separations, respectively. Analogous investigations for the other samples listed in Table \ref{tab:Geometry} are presented in the Supplemental Material \cite{Basaric-suppl2026}. The measurements were carried out in an axial magnetic field ranging between $-2$ and 2\,T, and under varying gate voltages $V_\mathrm{g}$ ranging between $-3$ and 3\,V, in 0.1\,V steps. In Figs.~\ref{fig:Figure-Gate} (a) and (b), the normalized magnetoconductance $G/G_0$ for samples C and E, with $G_0=e^2/h$, reveals regular AB-type oscillations. These oscillations span the entire gate voltage range, and are consistent even upon applying a gate voltage up to $-3$\,V, implying preservation of phase-coherent transport and a persistent ring geometry of the electron wave function upon subtle carrier depletion.

The ring-shaped wave function encloses an effective area $S$, which in case of our nanowires corresponds to a hexagonal cross section $S=r^{2}(3\sqrt{3}/2)$, where $r$ is the circumradius of the hexagon. The oscillation period $\Delta B$ of the AB type oscillations is associated to the nanowire cross section $S$ via the expression $\Delta B= \Phi_0/S$, where $\Phi_0=h/e$ is the magnetic flux quantum. For samples C and E, the extracted oscillation periods $\Delta B$ of 0.33 and 0.32\,T, imply a radius $r_{h/e}$ of the closed-loop wave function of around 69 and 70\,nm, respectively. This confirms that the wave function is localized within the InAs shell, and is found to be the case for all the other measured samples (see Table \ref{tab:Geometry}). Furthermore, sample E showcases smaller average conductance when compared to sample C, due to larger contact separation length.

Next, the fast Fourier transform (FFT) of the magnetoconductance oscillations was analyzed. The FFT was applied to the extracted oscillation signal $\delta G=G-\overline{G}$, where $\overline{G}$ is the slowly varying background which was subtracted from the measured conductance. The results for samples C and E presented in Figs.~\ref{fig:Figure-Gate} (c) and (d) show a clear two-peak structure over the entire gate voltage range. The first peak is associated with AB oscillations with an $h/e$ periodicity, and the second peak corresponds to oscillations with an $h/2e$ periodicity. The latter can originate from either the second harmonics of the AB oscillations or the AAS-type oscillations. Additionally, we find indications of higher harmonics for both samples, i.e. 3$^\mathrm{rd}$ and 4$^\mathrm{th}$ harmonics, though with significantly smaller amplitude. However, these higher harmonics are not present consistently throughout the complete gate voltage range. In Table~\ref{tab:Geometry}, the extracted positions of these peaks imply the presence of $h/3e$-, and $h/4e$-periodic oscillations. The slight broadening of the FFT peaks observed for sample E is attributed to its larger measured nanowire segment. In this case, a larger number of phase-coherent trajectories with small radius variations contribute to a broader recorded frequency spectrum.

Higher harmonics of AB oscillations have already been observed in other types of structures such as graphene \cite{DauberStampfer17}, or two-dimensional electron gas heterostructure GaAs/Ga$_x$Al$_{1-x}$As rings \cite{Hansen01}, as well as in Bi$_2$Se$_3$ topological insulator nanowires \cite{Dufouleur13}. In these cases, the occurrence of higher harmonics stems from the system's very long phase coherence length, which allows electron partial waves to encircle the ring multiple times. However, this phenomenon has not yet been reported for GaAs/InAs core/shell nanowires. Its transport in terms of angular momentum states does not provide a straightforward explanation that aligns with the interferometric, trajectory-based framework outlined in previous studies.

To further understand the underlying physical origin of our observations, we present a quantitative study of the phase rigidity of AB and AAS oscillations and their respective higher harmonics, through varying inner contact separation lengths. For that purpuse we extracted the frequency windows characteristic of the specific harmonics of the conductance oscillation (cf. Table~\ref{tab:Geometry}), and displayed their corresponding oscillation amplitude in a colormap as a function of magnetic field and gate voltage. Following this method, the analysis of $h/e$-periodic oscillations for sample C (cf. Fig.~\ref{fig:Figure-Gate} (e)) reveals a nonrigid phase pattern at $B=0$ as the gate voltage varies. The random phase shift along zero magnetic field is clearly confirmed by the solid blue line in Fig.~\ref{fig:Figure-Gate} (i), which illustrates the diminished averaged conductance over the voltage range $\langle \delta G \rangle_{V_\mathrm{g}}$ at $B=0$, whereas featuring large standard deviation marked in green at the same time. Sample E shows a similar behavior, as can be seen in Figs.~\ref{fig:Figure-Gate} (m) and (q). 

Carrying out an analogous analysis for the $h/2e$-periodic oscillations, shown in Fig.~\ref{fig:Figure-Gate} (f), sample C reveals a phase-rigid behavior in the low magnetic field range, with a conductance maximum at $B=0$. This observation is further supported through the analysis of $h/2e$-periodic conductance oscillations averaged over the gate voltage range (cf. Fig.~\ref{fig:Figure-Gate} (j)). Here, the phase stability results in a pronounced oscillatory behavior of the averaged value extending over a magnetic field range of about $\pm\,0.4\,$T, with its maximum value at $B=0$. Additionally, the small standard deviation further confirms the observed phase rigidity. The phase-rigid part is associated with AAS conductance oscillations, as such oscillations are robust against averaging over the $V_\mathrm{g}$ range. The source of these oscillations is the interference of time-reversed paths in the InAs shell of our tubular conductor. The presence of a conductance maximum at $B=0$ implies the presence of spin-orbit coupling via the weak antilocalization effect \cite{Aronov1987}. A similar pattern is found for sample E, as shown in Figs.~\ref{fig:Figure-Gate} (n) and (r). Here, the phase-rigid segment also extends to about $\pm 0.4\,$T. By comparing the phase rigidity expressed by the standard deviation, it can be concluded that the phase-rigid region of sample E is more robust against averaging over the entire gate voltage range than sample C. Beyond this range, the magnetoconductance oscillations remain to contain an $h/2e$-periodic component, but exhibit random phase shift due to the loss of phase matching along the inner and outer radius of InAs shell \cite{Mur2008}. We associate displayed $h/2e$-periodic oscillations beyond $\pm\,0.4\,$T to the second harmonic of AB oscillations, which may also contribute to a small extent around zero field. The phenomena associated with $h/2e$ oscillations have already been addressed in our recent study of quasi-ballistic transport properties in phase-pure GaAs/InAs nanowires \cite{Basaric2025}. In addition to the data presented for samples C and E, $h/2e$-periodic oscillations were also observed for the other samples listed in Table \ref{tab:Geometry} (see Supplemental Material \cite{Basaric-suppl2026}). 
 
\begin{figure}[!h]
    \centering 
    \includegraphics[width=\columnwidth]{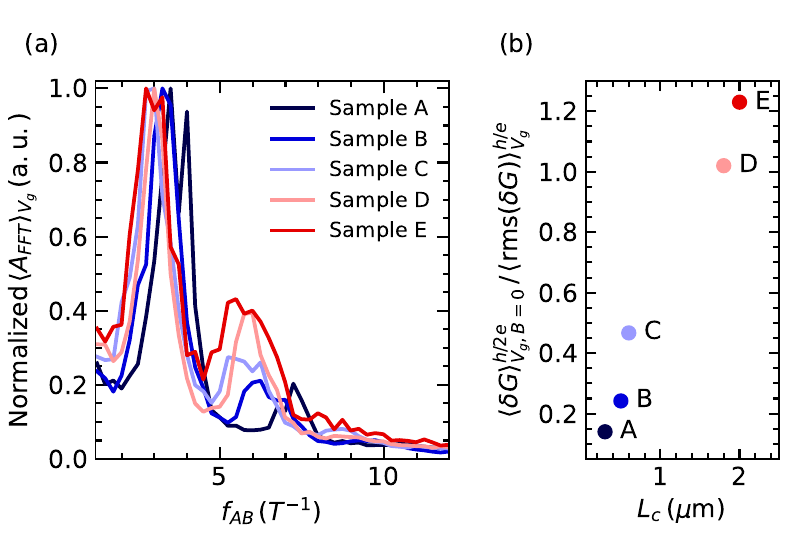}
    \caption{\justifying(a) Normalized average of fast Fourier Amplitude $\langle A_\mathrm{FFT} \rangle_{V_g}$ as a function of frequency for samples with different contact separations (samples A to E). (b) Ratio of the average amplitude of the AAS oscillations $\langle \delta G \rangle^{h/2e}_{V_\mathrm{g},B=0}$ at $B=0$ and the root-mean-square amplitude of the $h/e$-periodic AB-type oscillations, averaged over the gate voltage $\langle rms(\delta G)\rangle^{h/2}_{V_\mathrm{g}}$ for different contact separation lengths.} 
    \captionsetup{position=bottom}
    \label{fig:Figure-analysis}
\end{figure}

The FFT analysis of the magnetoconductance of samples C and E contain $h/3e$- and $h/4e$-periodic oscillations. As can be seen in the colormap of the $h/3e$-periodic oscillations of sample C in Fig.~\ref{fig:Figure-Gate} (g), the conductance predominantly exhibits a maximum at $B=0$ with some short sections of a minimum. Similar to the $h/2e$-periodic oscillations the average value plotted in Fig.~\ref{fig:Figure-Gate} (k) shows a clear oscillation pattern with a maximum at $B=0$, which is damped out with increasing field. The $h/3e$-periodic oscillations of Sample E shown in Fig.~\ref{fig:Figure-Gate} (o) reveals an even more stable phase-rigid behavior compared to sample C, with a gradual oscillation amplitude decay with carrier depletion at larger negative gate voltages. As shown in Fig.~\ref{fig:Figure-Gate} (s) the standard deviation is small in this magnetic field range and also smaller than the standard deviation for sample C given in Fig.~\ref{fig:Figure-Gate} (k). We assume that the phase rigidity of the $h/3e$-periodic oscillations can be attributed to the propagation of the phase rigidity of the $h/2e$-periodic oscillations, under the premise that the AAS-type oscillations are damped out with an increasing magnetic field, as it is actually observed in Figs.~\ref{fig:Figure-Gate} (j) and (r). Unlike the phase rigidity of the $h/2e$-periodic oscillations, originating from the AAS-effect, the phase rigidity of the $h/3e$-periodic oscillations does not correspond to a distinct physical phenomenon. 

For the amplitude of the $h/4e$-periodic oscillations of sample C, the color map shown in Fig.~\ref{fig:Figure-Gate} (h) reveals a pronounced phase-rigid region approximately half the size of the corresponding $h/2e$-periodic oscillation, and a small standard deviation, shown in Fig.~\ref{fig:Figure-Gate} (l). Beyond this magnetic field range, the regular phase shifts along the gate voltage range, indicating the previously described phase randomization. As can be inferred from Fig.~\ref{fig:Figure-Gate} (p), once again for sample E the phase rigidity is more pronounced, supported by the even smaller standard deviation at $B=0$, shown in Fig.~\ref{fig:Figure-Gate} (t). The phase rigidity of the $h/4e$-periodic oscillations can be understood within the framework of the AAS-effect when considering quasi-classical trajectories that circumvent the tubular conductor twice for each time-reversed paths thus effectively encircling the magnetic flux four times. This interpretation is supported by the fact that the phase-rigid magnetic field range for the $h/4e$-periodic oscillations is approximately half that for the $h/2e$-periodic oscillations, as seen by comparing Figs.~\ref{fig:Figure-Gate} (l) and (t) with (j) and (r), which can be explained by the larger phase difference accumulated for inner and out trajectories when circumventing twice. However, our experimental findings of $h/3e$-, and $h/4e$-periodic oscillations still remain in need of further analysis. A detailed discussion of results gained through tight-binding model simulations is presented in the following section.
 
As only three samples of the ones listed in Table \ref{tab:Geometry} display $h/3e$- and $h/4e$-periodic oscillations, we will limit our understanding of overall tendencies of phase-coherent transport with varying inner contact lengths through more detailed analysis of $h/e$- and $h/2e$-periodic oscillations. First, comparing the oscillations shown in Figs.~\ref{fig:Figure-Gate} (e) and (m), we observe a tendency for $h/e$-periodic oscillations amplitude reduction for increasing $L_c$. This is attributed to an increasing amount of averaging out of non-phase rigid AB-type oscillations along the nanowire. In Fig.~\ref{fig:Figure-analysis} (a), the ensemble averaged normalized FFT spectrum $\langle A_\mathrm{FFT} \rangle_{V_g}$ is shown for samples with different contact separation lengths $L_\mathrm{c}$. Here, we observe a significant reduction of $h/2e$-periodic oscillation FFT amplitude with increasing $L_c$. However, since the FFT spectra were taken over the entire magnetic field range of $\pm2$\,T, the $h/2e$-periodic oscillations can originate from the second harmonics of the AB-type oscillations and from the AAS effect. As disclosed earlier in the text, the $h/2e$-periodic oscillations contain a phase-rigid section around $B=0$, and non-phase rigid section beyond. To properly discriminate the relationship between the AB- and AAS-effect, the ratio of the average amplitude of the AAS-type oscillations $\langle \delta G \rangle^{h/2e}_{V_\mathrm{g},B=0}$ at $B=0$, and the root-mean-square amplitude of the $h/e$-periodic AB-type oscillations averaged over the gate voltage $\langle rms(\delta G)\rangle^{h/2e}_{V_\mathrm{g}}$ is shown in Fig.~\ref{fig:Figure-analysis} (b) for varying $L_c$. It is evident that the ratio increases with increasing contact separation length, indicating an increasing averaging out of the AB-effect and a simultaneous enhancement of the phase-rigid AAS contribution. This interpretation is further supported by data on the averaged oscillation amplitude and the extracted FFT peak height of the $h/e$ component given in the Supplemental Material \cite{Basaric-suppl2026}. 

\section{Quantum Transport Simulations}

To gain a deeper understanding of the interrelation of AB- and AAS-type oscillations for different nanowire lengths, quantum transport simulations were conducted. Details on the simulation method and parameters are provided in the Supplemental Material \cite{Basaric-suppl2026}. The calculations were performed using the KWANT software package \cite{Groth2014}. The assumed radial dimensions of the hexagonally shaped core/shell nanowires were comparable to the experimental ones. Disorder in the wires was described by a Gaussian random field with a disorder strength of 5\,meV and a correlation length of 50\,nm. Both, Rashba spin-orbit coupling and Zeeman splitting were included. Instead of the gate voltage, for the simulations the chemical potential $\mu$ was varied from 10 to 50\,meV. The simulations were performed for a temperature of 1.6\,K.  

As an initial test, we performed simulations on a 2-\textmu m-long nanowire with and without the presence of spin-orbit coupling. Here, we could observe a clear minimum of conductance at $B=0$ in the former case, while a maximum was found in the latter case, in accordance with the weak localization and weak antilocalization effects, respectively. More details on this simulation aspect are given in the Supplemental Material \cite{Basaric-suppl2026}. 

\begin{figure*}[!htbp]
\centering
\includegraphics[width=\textwidth]{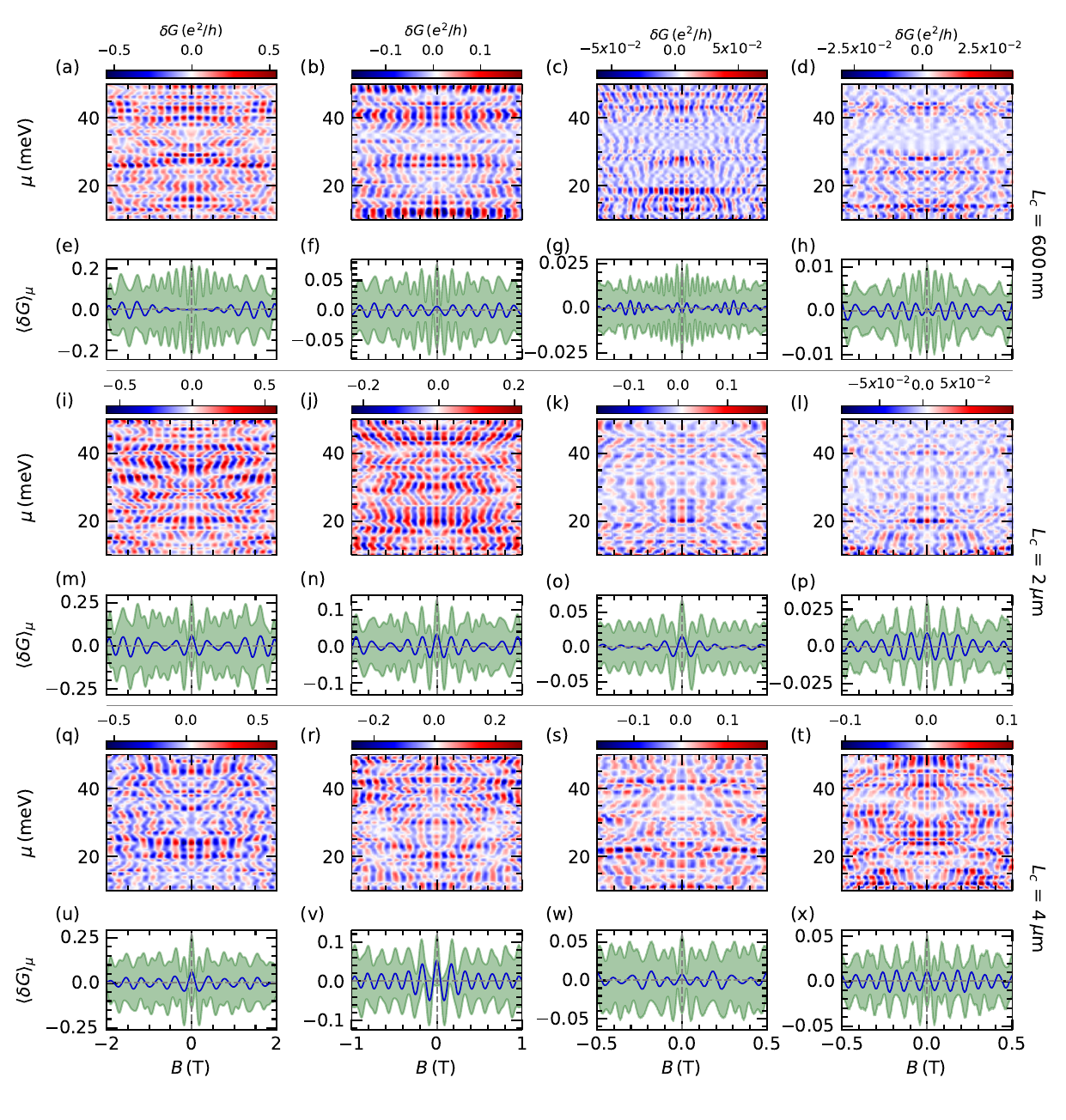}
\caption{\justifying Panels (a)-(d), (i)-(l), and (q)-(t) show the conductance oscillation amplitudes of first to fourth harmonics, with periods from $h/e$ to $h/4e$, as a function of the magnetic field and chemical potential, for nanowires of lengths $L=600\,$nm, $2\,${\textmu}m, and $4\,${\textmu}m, respectively. Panels (e)-(h), (m)-(p), and (u)-(x) display the average amplitude of the oscillation (solid blue line) and its standard deviation (in green) with respect to the chemical potential $\mu$ for each length. Note that the displayed magnetic field range is successively reduced for the higher harmonics.}
\label{fig:simulation}
\end{figure*}

Figure~\ref{fig:simulation} summarizes the simulation results for three different nanowire lengths: 0.6, 2, and 4\,{\textmu}m. The Fourier components related to the periods $h/e$, $h/2e$, $h/3e$, and $h/4e$ were extracted from the total conductance, interpolated and plotted as a function of $B$ and $\mu$. Additionally, the corresponding conductance averaged over $\mu$, denoted as $\langle \delta G \rangle_\mu$, and its standard deviation are shown. For all wire lengths, we observe pronounced regular oscillations of the $h/e$-periodic components of the conductance, with phase shifts as $\mu$ is varied, as shown in Figs.~\ref{fig:simulation} (a), (i), and (q). At $B=0$ the phase switches repeatedly with $\mu$. In our simulations, maxima in $\delta G$ appear when the density of state possesses a maximum at $\mu$. Standard deviation $\langle \delta G \rangle_\mu$ marked in green for three wire lengths in Figs.~\ref{fig:simulation} (e), (m), and (u) at $B=0$ implies a positive phase bias, which we attribute to the particular impurity configuration used in our simulation. The $h/2e$-periodic conductance component shown in Figs.~\ref{fig:simulation} (b), (j), and (r), however, exhibits an almost continuous maximum at zero magnetic field for large intervals of the chemical potentials. As also observed in the experimental data, the phase-rigid feature persists until larger critical magnetic field with the length of the nanowire, extending over a few units of magnetic flux quantum. Additionally, decrease of standard deviation $\langle \delta G \rangle_\mu$ at $B=0$ with nanowire length increase, displayed in Figs.~\ref{fig:simulation} (f), (n), and (v), aligns well with our experimental findings given in Figure~\ref{fig:Figure-analysis} (b).

In Figs.~\ref{fig:simulation} (c) and (d), extracted oscillations amplitudes for $h/3e$- and $h/4e$-periodic components of the 600$\,$nm-long nanowire are depicted. However, the extracted average conductance given in Figs.~\ref{fig:simulation} (g) and (h) give a better qualitative overview, as it showcases a small minimum at zero magnetic field. Nevertheless, the standard deviation is rather large, therefore we interpret this as an accidental occurrence. In contrast to this, Figs.~\ref{fig:simulation} (k) and (l), as well as Figs.~\ref{fig:simulation} (s) and (t), for the 2\,{\textmu}m and 4\,{\textmu}m long nanowires exhibit a tendency towards a rigid positive phase at $B=0$ for both, $h/3e$- and $h/4e$-periodic components. Averaged conductance in case of both components presented in Figs.~\ref{fig:simulation} (o) and (p) for 2\,{\textmu}m, and Figs.~\ref{fig:simulation} (w) and (x) for 4\,{\textmu}m long nanowires, seems to decrease with increasing nanowire length. However, in case of the $h/3e$-periodic oscillations the standard deviation is relatively large, so the phase rigidity is not as robust as in case of the $h/4e$-periodic oscillations. The latter is not surprising since the $h/4e$-periodic oscillations contain the second harmonics of the phase rigid $h/2e$-periodic AAS-type oscillations. In addition to that, the phase rigidity of the AAS-type oscillations might also have propagated into the first and third harmonics. This claim is supported by calculation results presented in Supplemental Material \cite{Basaric-suppl2026}.

\begin{figure}[!htbp]
\centering
\includegraphics[width=\columnwidth]{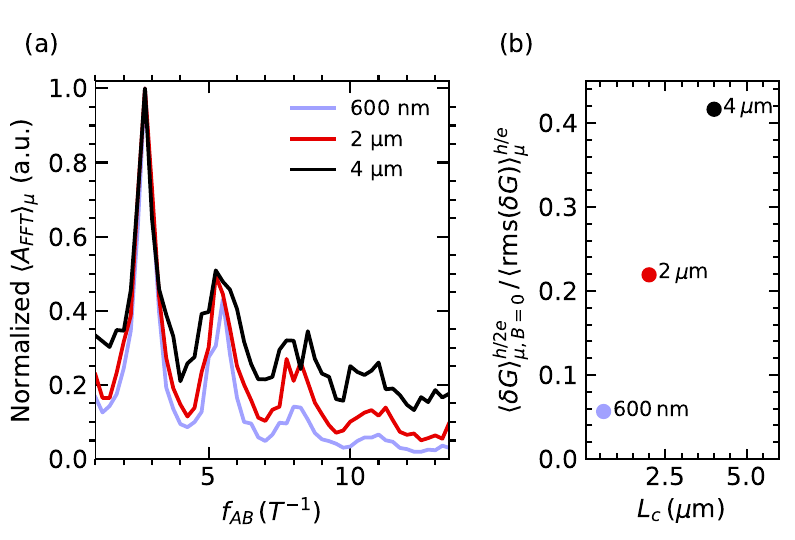}
\caption{\justifying (a) Normalized fast Fourier amplitude $\langle A_\mathrm{FFT} \rangle_\mu$ averaged over $\mu$ as a function of frequency $f$ for the 600\,nm (light blue), 2\,{\textmu}m (red), and 4\,{\textmu}m (dark blue) long nanowires. (b) Ratio of the average amplitude of the simulated AAS-type oscillations $\langle \delta G \rangle^{h/2e}_{\mu,B=0}$ at $B=0$ and the root-mean-square amplitude of the simulated $h/e$-periodic AB-type oscillations averaged over the chemical potential $\langle rms(\delta G)\rangle^{h/2e}_{\mu}$.}
\label{fig:AveragedFFT}
\end{figure}

Analogously to the analysis of FFT and oscillation amplitude dependency to varying inner contact length depicted in Fig.~\ref{fig:Figure-analysis}, we will take a closer look at computed conductance oscillations and its contributions dependencies to varying nanowire length. In Fig.~\ref{fig:AveragedFFT} (a) the ensemble averaged normalized FFT spectra over the chemical potential is shown for the 600\,nm, 2\,{\textmu}m, and  4\,{\textmu}m long nanowires (the single spectra for each $\mu$ are given in the Supplemental Material \cite{Basaric-suppl2026}). Peaks corresponding to the $h/e$-, $h/2e$-, $h/3e$-, and $h/4e$-periodic oscillations are clearly resolved, with decreasing peak heights towards higher frequencies. Consistent with the experimental results, we observe that the peak corresponding to the $h/2e$-periodic oscillations is more pronounced for the longer nanowire, confirming the increasingly larger AAS contribution with increasing nanowire length. Finally, Fig.~\ref{fig:AveragedFFT} (b) depicts the ratio of the average amplitude of the simulated AAS-type oscillations $\langle \delta G \rangle^{h/2e}_{\mu,B=0}$ at $B=0$ and the root-mean-square amplitude of the $h/e$-periodic AB-type oscillations averaged over the chemical potential $\langle rms(\delta G)\rangle^{h/2e}_{\mu}$ for the 600\,nm, 2\,{\textmu}m, and  4\,{\textmu}m long nanowires. In accordance with the experimental results shown in Fig.~\ref{fig:Figure-analysis} (b), the ratio clearly increases with increasing nanowire length, caused by an increasing averaging out of the AB-effect and an enhanced AAS-contribution. 

\section{Conclusions}

We systematically studied flux-periodic oscillations of GaAs/InAs core/shell nanowires at the fundamental period of $h/e$ up to its fourth harmonic $h/4e$. The $h/e$-periodic oscillations are attributed to the Aharonov--Bohm effect, which is non-phase-rigid for different phase-coherent segments of the nanowire, or when the gate voltage is changed. Nevertheless, even for the longest contact separation length of 2\,\textmu m, we found pronounced $h/e$-periodic oscillations, although with a smaller amplitude,  indicating weak disorder-related averaging. This is probably due to the high crystal quality of our phase-pure nanowires. In contrast to the averaged oscillation amplitude vanishing for $h/e$-periodic component, the $h/2e$ oscillations are phase rigid within a certain range around zero field. For these oscillations, phase rigidity increases with longer contact separation, likely due to the diminishing contribution of second harmonics of AB-type oscillations. The latter are visible at larger magnetic fields. Surprisingly, phase rigidity is also observed in the $h/3e$-periodic oscillations, which cannot be directly assigned to interference of time-reversed paths. Through theoretical analysis, we discovered that the phase rigidity of the $h/3e$-periodic oscillations is due to the phase rigidity of the lower harmonics, i.e. the $h/2e$-periodic AAS-type oscillations. We identified a robust phase-rigid magnetic field range for the $h/4e$-periodic oscillations and found that this range is about half as wide as the one of the $h/2e$-periodic oscillations. In general, we can conclude that, despite the short contact separation for some of our nanowires and limited disorder, we clearly observe Altshuler--Aronov--Spivak-type interference effects, which are phase-rigid and usually attributed to larger ensembles of time-reversed trajectories in the diffusive regime. Our transport simulations investigating the role of increasing wire length also support this conclusion. Another interesting aspect is the occurrence of higher harmonics up to the fourth one. This implies that the conductance oscillations originating from the Aharonov--Bohm effect and weak antilocalization effects deviate from pure sinusoidal dependence. This can be attributed to the quasi-ballistic transport regime associated with low disorder, which in the semiclassical trajectory picture allows interference of multiple winding paths. In the fully quantum-mechanical picture, one can interpret the presence of higher harmonics that features in the density of states of the nanowire are prevented to be averaged out.

\section*{Acknowledgements}

We thank Herbert Kertz for technical assistance. Dr. Florian Lentz and Dr. Stefan Trellenkamp are gratefully acknowledged for their help with the required e-beam lithography and Benjamin Bennemann and Christoph Krause for their contribution to the growth of the nanowires. We furthermore thank Dr. Kristof Moors  and William Schaarman for fruitful discussions and for providing the initial code for the KWANT simulations. The sample fabrication has been performed in the Helmholtz Nano Facility at Forschungszentrum J\"ulich \cite{Albrecht2017}. This work was partly funded by Deutsche Forschungsgemeinschaft (DFG, German Research Foundation) under Germany’s Excellence Strategy—Cluster of Excellence Matter and Light for Quantum Computing (ML4Q) EXC 2004/2—390534769. A.M.S. and F.A.C. acknowledge the support of the EPSRC Grant EP/W002418/1. T.G.D. was funded by Reykjavik University Research Fund, project 223016. F.B. and T.S. would like to express their gratitude for the support provided as part of the ASPIRE project JPMJAP2338.

\section*{Data Availability}

The data that support the findings of this article are openly
available  \cite{Basaric-data2026}.

\putbib[bu1.bbl]  
\end{bibunit}

\clearpage
\widetext

\titleformat{\section}[hang]{\bfseries}{\MakeUppercase{Supplemental Note} \thesection:\ }{0pt}{\MakeUppercase}
\begin{bibunit}[]
\setcounter{section}{0}
\setcounter{equation}{0}
\setcounter{figure}{0}
\setcounter{table}{0}
\setcounter{page}{1}
\renewcommand{\thesection}{\arabic{section}}
\renewcommand{\thesubsection}{\Alph{subsection}}
\renewcommand{\theequation}{S\arabic{equation}}
\renewcommand{\thefigure}{S\arabic{figure}}
\renewcommand{\figurename}{Supplemental Figure}
\renewcommand{\tablename}{Supplemental Table}
\renewcommand{\bibnumfmt}[1]{[S#1]}
\renewcommand{\citenumfont}[1]{S#1}

\begin{center}

\textbf{\large Supplemental Material: Interplay between Aharonov--Bohm and Altshuler--Aronov--Spivak oscillations in phase-pure GaAs/InAs core/shell nanowires of different lengths}
\end{center}

{
  \hypersetup{linkcolor=black}
  \tableofcontents
}


\section{Experimental Details}

The investigated core/shell GaAs/InAs nanowires were grown by molecular beam epitaxy (MBE) via self-catalyzed vapor-liquid-solid technique on pre-structured substrates. We prepared Si(111) substrates with about 20\,nm of thermally deposited $\mathrm{SiO}_{2}$, consisting of hole arrays with varying diameters of 40, 60, and 80\,nm and pitches with varying pinhole sizes of 0.5, 1, 2, and 4\,$\upmu$m. The arrays were defined by electron beam lithography, followed by dry and wet etching. For realization of a phase-pure crystal structure, the catalyst droplet on top of the growing nanowire was dynamically controlled, following the approach reported in Jansen \textit{et al.} \cite{Jansen2020}. In our case, a contact angle of the Ga catalyst droplet above $125^\circ$ was achieved, therefore the investigated core/shell nanowires contain a zinc-blende (ZB) crystal structure. First, the GaAs core was grown by applying an As flux with a beam equivalent pressure (BEP) of $5\times 10^{-6}$ mbar for 90\,min at about $615^\circ$C, during which the Ga flux was dynamically decreased by 40\,\% from the starting value of $1.5\times 10^{-7}$\,mbar, resulting in average diameter of $65\,$nm. Furthermore, the InAs shell growth was performed at a substrate temperature of $450^\circ$C, comprising In and As fluxes with BEPs of $3.0\times 10^{-7}$ mbar and $5\times 10^{-6}$ mbar, respectively, yielding average thickness of $30\,$nm. Table 1 in the main text presents the resulting nanowire geometries, where slight dimension variations appear as the nanowires originate from different pitches.

\section{Transmission Electron Microscopy Studies}

\begin{figure*}[tbh]
    \centering 
     \includegraphics[width=0.95\textwidth]{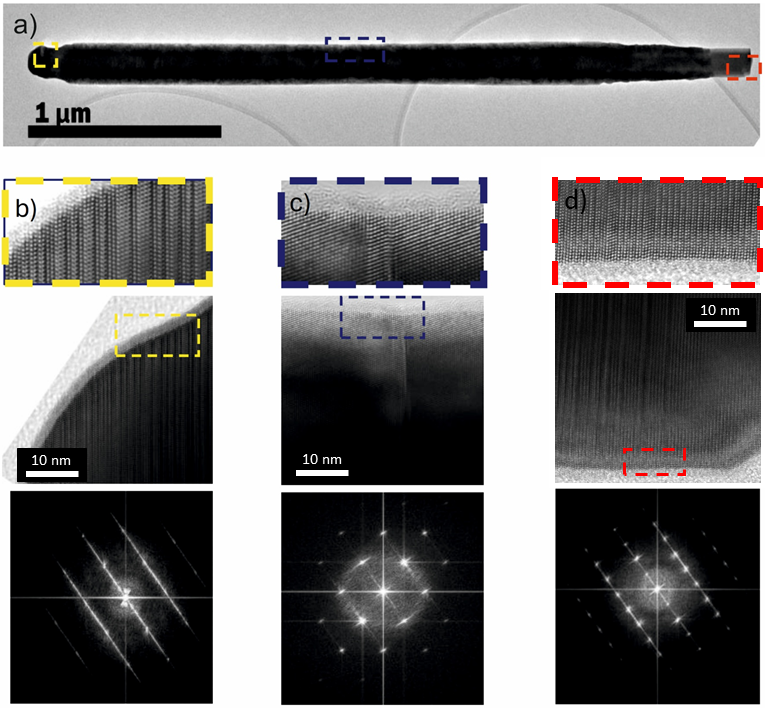}
    \caption{\justifying(a) Bright-field image of the nanowire, with length of 3.8\,{\textmu}m and diameter of 213\,nm in the middle part. (b) High-resolution TEM image corresponding to the yellow box in (a), showing stacking faults and different polytypes. (c) Middle part of the nanowire, exhibiting a preferred zincblende structure. (d) Region displaying twins and stacking faults with both wurtzite and zincblende phases.} 
    \captionsetup{position=bottom}
    \label{fig:Figure-TEM-supp}
\end{figure*}

The nanowires crystal structure was analysed by transmission electron microscopy (TEM). Here, nanowires from the same growth run as the ones characterized in transport experiments were transferred onto a lacey carbon grid, and analyzed using a JEOL2100 TEM operating at 200\,kV. The images were acquired along the $<$110$>$  zone axis. From Fig.~\ref{fig:Figure-TEM-supp} (a), the nanowire length is estimated to be approximately 3.8\,{\textmu}m, with a diameter of about 213\,nm at its midpoint.
High resolution images of the regions marked with yellow, blue and red rectangles in Figures~\ref{fig:Figure-TEM-supp} (b) - (d) provide a closer view of the nanowire structure. In Fig.~\ref{fig:Figure-TEM-supp} (b), corresponding to the yellow rectangle of (a), a polytypic structure can be distinguished, with twins and stacking faults, with both zincblende and wurtzite phases, further confirmed by the FFT pattern. Figure~\ref{fig:Figure-TEM-supp} (c) represents  the middle portion of the nanowire, where the zincblende crystal structure is the preferred polytype, as supported by the FFT shown below, which reveals the characteristic spots of zincblende.  Finally,  Fig.~\ref{fig:Figure-TEM-supp} (d), corresponding to the red rectangle in (a), highlights the presence of stacking faults, twinning and wurtzite structure. 

\section{$h/e$ and higher harmonics conductance contributions for all samples}

\begin{figure*}[tbh]
    \centering 
     \includegraphics[width=0.65\textwidth]{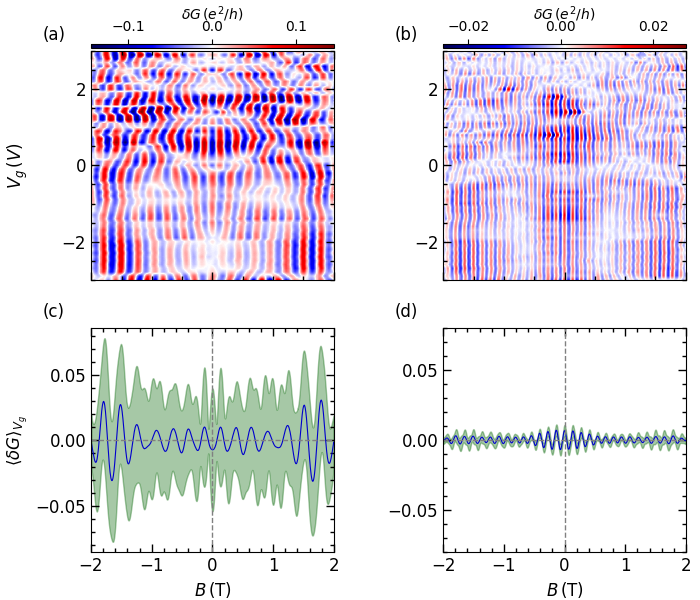}
     \caption{\justifying Sample A ($L_\mathrm{c}=300$\,nm): (a),(b) Color maps of the filtered-out conductance oscillations as a function of magnetic field $B$ and gate voltage $V_\mathrm{g}$ for the $h/e$- and $h/2e$-periodic components, respectively. (c),(d) Conductance oscillations averaged over the gate voltage as a function of magnetic field and corresponding standard deviation for the $h/e$ and $h/2e$ periodic components, respectively.}
    \captionsetup{position=bottom}
    \label{fig:Figure-SampleA-all}
\end{figure*}

\begin{figure*}[tbh]
    \centering 
     \includegraphics[width=0.65\textwidth]{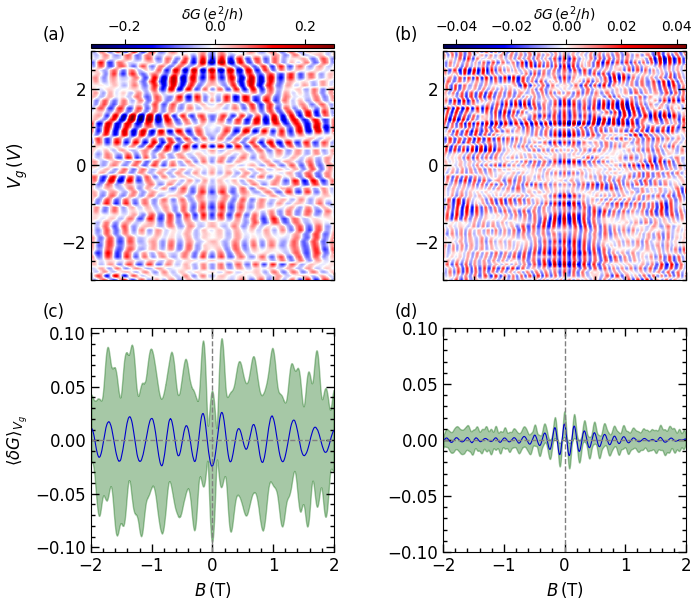}
     \caption{\justifying Sample B ($L_\mathrm{c}=500$\,nm): (a) - (d) Color maps of the filtered-out conductance oscillations as a function of magnetic field and gate voltage for the $h/e$-, $h/2e$-, $h/3e$-, and $h/4e$-periodic components, respectively. (e) - (h) Conductance oscillations averaged over the gate voltage as a function of magnetic field and corresponding standard deviation for the $h/e$-, $h/2e$-, $h/3e$-, and $h/4e$-periodic components, respectively.}
    \captionsetup{position=bottom}
    \label{fig:Figure-SampleB-all}
\end{figure*}

\begin{figure*}[tbh]
    \centering 
     \includegraphics[width=\textwidth]{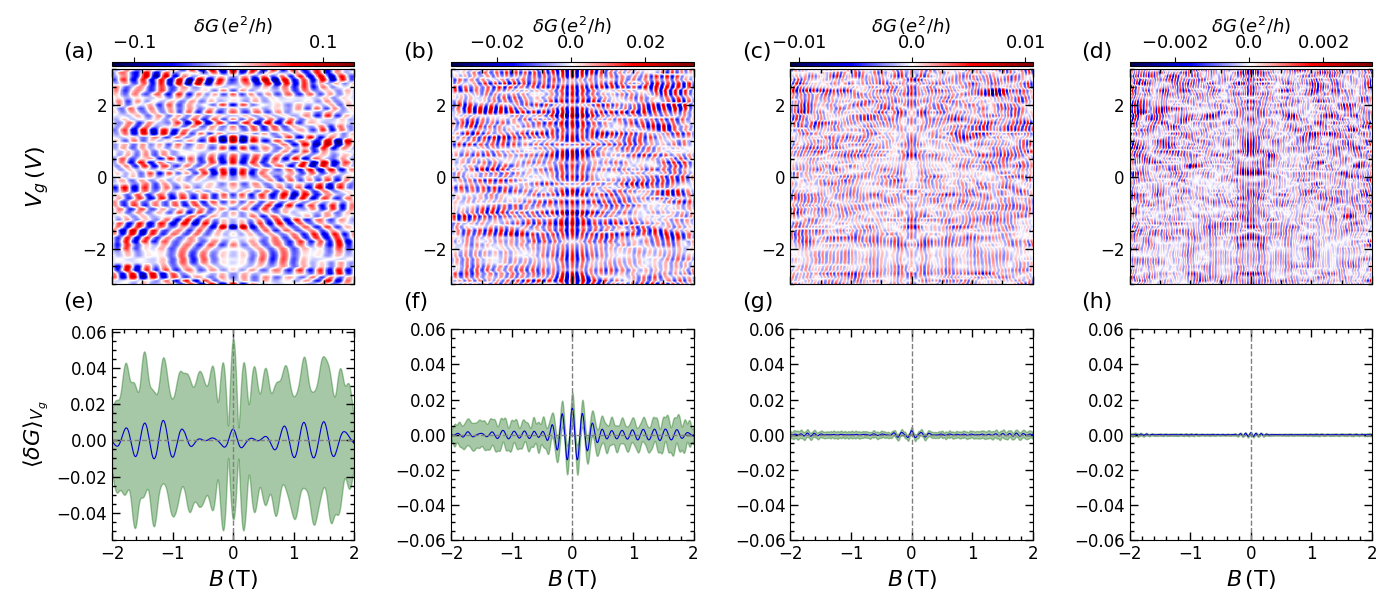}
     \caption{\justifying Sample C ($L_\mathrm{c}=600$\,nm): (a) - (d) Color maps of the filtered-out conductance oscillations as a function of magnetic field and gate voltage for the $h/e$-, $h/2e$-, $h/3e$-, and $h/4e$-periodic components, respectively. (e) - (h) Conductance oscillations averaged over the gate voltage as a function of magnetic field and corresponding standard deviation for the $h/e$-, $h/2e$-, $h/3e$-, and $h/4e$-periodic components, respectively.}
    \captionsetup{position=bottom}
    \label{fig:Figure-SampleC-all}
\end{figure*}

\begin{figure*}[tbh]
    \centering 
     \includegraphics[width=\textwidth]{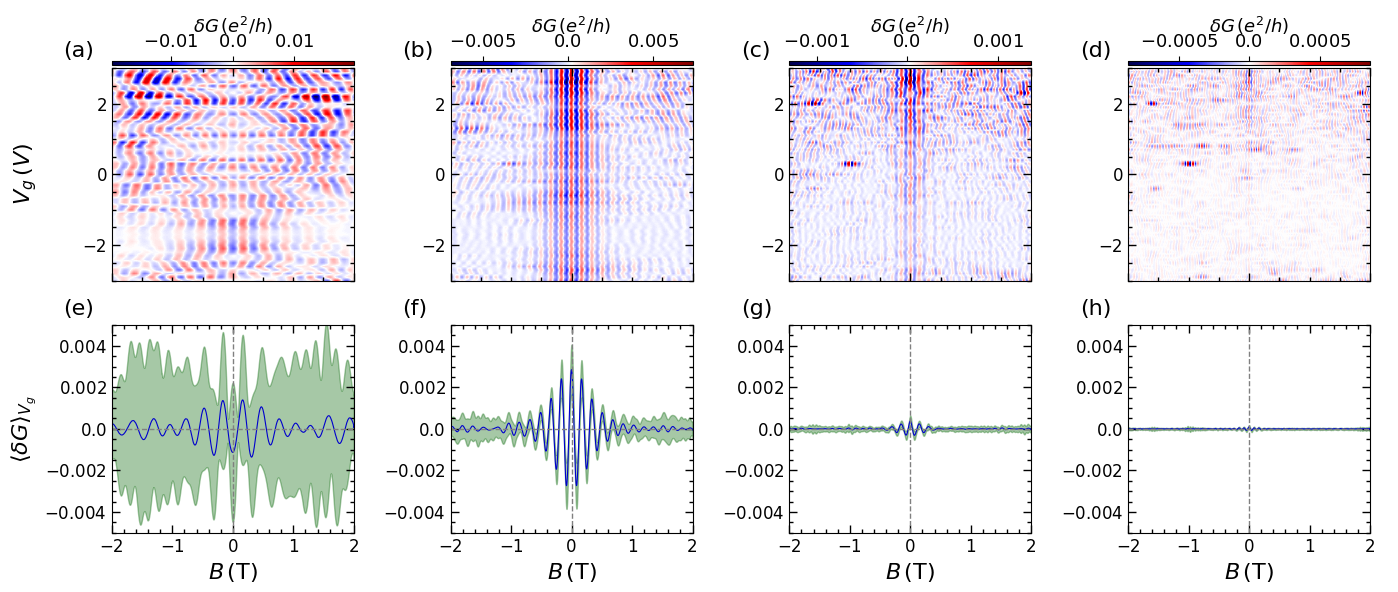}
     \caption{\justifying Sample D ($L_\mathrm{c}=1800$\,nm): (a) - (d) Color maps of the filtered-out conductance oscillations as a function of magnetic field and gate voltage for the $h/e$-, $h/2e$-, $h/3e$-, and $h/4e$-periodic components, respectively. (e) - (h) Conductance oscillations averaged over the gate voltage as a function of magnetic field and corresponding standard deviation for the $h/e$-, $h/2e$-, $h/3e$-, and $h/4e$-periodic components, respectively.}
    \label{fig:Figure-SampleD-all}
\end{figure*}

\begin{figure*}[tbh]
    \centering 
     \includegraphics[width=\textwidth]{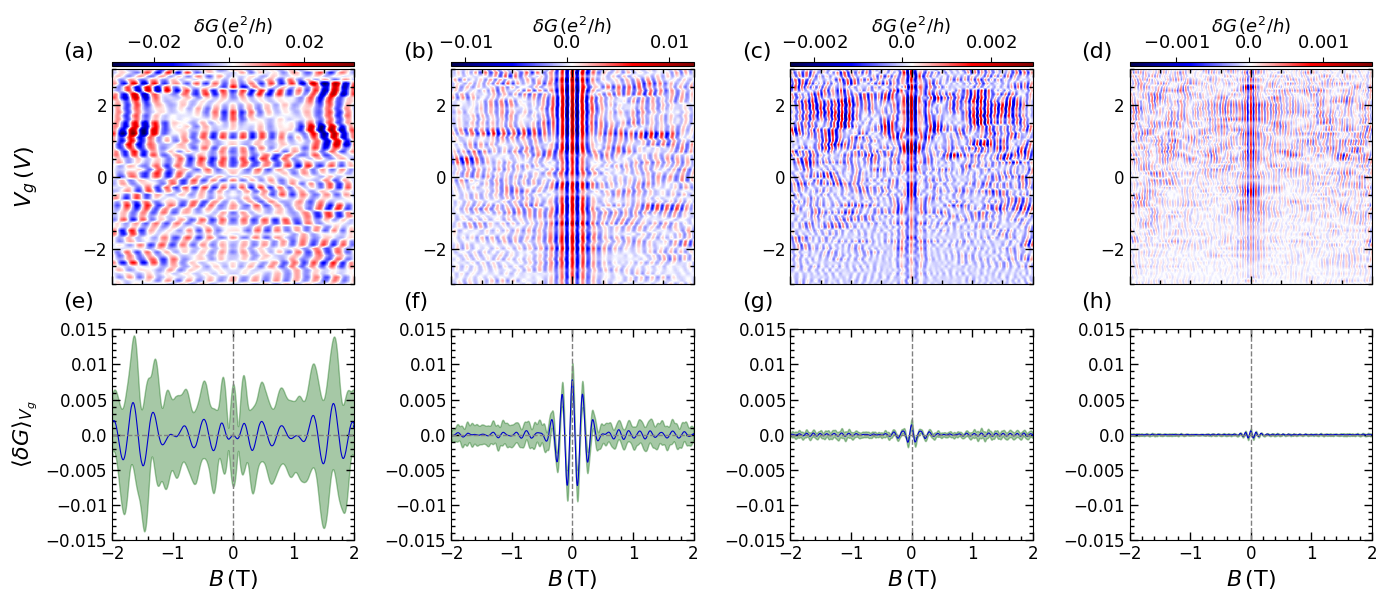}
     \caption{\justifying Sample E, ($L_\mathrm{c}=2000$\,nm): (a) - (d) Color maps of the filtered-out conductance oscillations as a function of magnetic field and gate voltage for the $h/e$-, $h/2e$-, $h/3e$-, and $h/4e$-periodic components, respectively. (e) - (h) Conductance oscillations averaged over the gate voltage as a function of magnetic field and corresponding standard deviation for the $h/e$-, $h/2e$-, $h/3e$-, and $h/4e$-periodic components, respectively.}
    \label{fig:Figure-SampleE-all}
\end{figure*}

Figures~\ref{fig:Figure-SampleA-all} to \ref{fig:Figure-SampleE-all} summarize the experimental data for the five samples under investigation. In Figs.~\ref{fig:Figure-SampleA-all} (a) and (b) the color maps show the filtered-out $h/e$- and $h/2e$-periodic conductance contributions, respectively, as a function of $B$ and $V_\mathrm{g}$ of sample A, which has the shortest contact separation of $L_\mathrm{c}= 300$\,nm. It has to be noted that this sample was measured in three-terminal setting, due to the fact that one contact was not working. However, approximate value of contact resistance of 1\,k$\Omega$ was subtracted from the measured signal in case of displayed plots. The $h/e$ contribution is apparently non-phase-rigid (cf. Figs.~~\ref{fig:Figure-SampleA-all} (a) and (c)). It is attributed to the Aharonov--Bohm (AB) effect. The $h/2e$ contribution shown in Fig.~\ref{fig:Figure-SampleA-all} (b) is rather weak but phase-rigid around zero field between $\pm 0.4$\,T due to the Altshuler--Aronov-Spivak (AAS) effect. This is confirmed in Fig.~\ref{fig:Figure-SampleA-all} (d), where the averaged conductance shows a clear oscillation signal with a small standard deviation. We attribute the relatively weak AAS contribution to averaging over only a few time-reversed trajectories due to the short contact separation length. The $h/2e$-periodic oscillations beyond the range of $\pm 0.4$\,T  are assigned to the non-phase-rigid second harmonics of the AB effect. For sample A, no $h/3e$- and $h/4e$-periodic contributions are observed. We also attribute this to the short contact separation length.    


For sample B with a contact separation length of 500\,nm we find a well developed AB and AAB effect but no substantial $h/3e$- and $h/4e$-periodic contributions, as can be seen in Fig.~\ref{fig:Figure-SampleB-all}. This is changed for samples C to E, where pronounced $h/3e$- and $h/4e$-periodic contributions are present, as can be found in Figs.~\ref{fig:Figure-SampleC-all} to \ref{fig:Figure-SampleE-all}, respectively.  

\clearpage
\newpage

\section{Extracted oscillation amplitude}

\begin{figure*}[tbh]
    \centering 
     \includegraphics[width=0.65\textwidth]{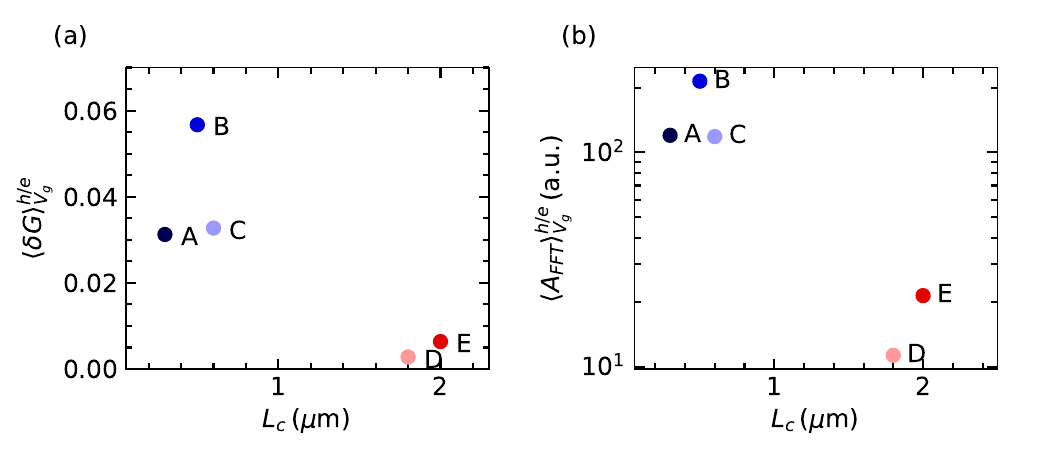}
    \caption{\justifying(a) Average of extracted $h/e$-periodic contribution $\langle \delta G \rangle_{V_g}^{h/e}$ over applied gate voltage for varying contact separation length $L_{c}$. (b) Averaged amplitude of FFT peak corresponding to a period of $h/e$ over applied gate voltage for varying $L_{c}$.} 
    \captionsetup{position=bottom}
    \label{fig:Figure-analysis-supp}
\end{figure*}

Figure~\ref{fig:Figure-analysis-supp} (a) shows the root-mean-square of the oscillation signal $\delta G=G-\overline{G}$, where $\overline{G}$ is the slowly varying background, as a function of contact separation length $L_\mathrm{c}$. As $L_c$ increases, the oscillation amplitude decreases, which can be explained by the averaging out of AB-type oscillations of different phases along the nanowire. These results align with the averaged amplitude of the peak corresponding to the $h/e$ period in the FFT spectrum for different $L_c$, shown in Fig.~\ref{fig:Figure-analysis-supp} (b).

\section{The model}

We used a quantum mechanical model of the GaAs/InAs core-shell nanowire to examine the conductance and its oscillations with respect to the axial magnetic field, in order to find out how the length of the nanowire might affect the magnitude of the higher harmonics and their phase rigidity at low magnetic fields. We model the nanowire, with its axis along the $x$-direction, using the following Hamiltonian:
\begin{equation}
\begin{alignedat}{2}
H_{\mathrm{sys}} \;&=\;
\left[
\frac{\hbar^{2}}{2m_{\mathrm{eff}}}\left(k_x^{2}+k_y^{2}+k_z^{2}\right)
+V_{\mathrm{sys}}\!\left(x,y,z\right)
\right] \mathbf{I} 
\\[4pt]
&-\alpha\Bigg[
k_x \Bigg(\sigma_{z}\sin\left(\theta_{\mathrm{so}}(y,z)\right)
+\sigma_{y}\cos\left(\theta_{\mathrm{so}}(y,z)\right)\Bigg)
-\sigma_{x}\Bigg(k_y\cos \left(\theta_{\mathrm{so}}(y,z)\right)
+k_z\sin\left(\theta_{\mathrm{so}}(y,z)\right)\Bigg)
\Bigg]
\\[4pt]
&+\frac{1}{2}\,g\,\mu_{B}\left(B_x\sigma_{x}+B_y\sigma_{y}+B_z\sigma_{z}\right) \; ,
\end{alignedat}
\label{eq:Hsys}
\end{equation}
which contains three distinct contributions. The first term describes the spin-independent single-particle motion in the scattering region, with wave vector $(k_x,k_y,k_z)$ and effective mass $m_{\mathrm{eff}}$, including the kinetic energy and the local potential $V_{\mathrm{sys}}$.  This potential incorporates the geometry of the nanowire and the internal disorder. The second term accounts for Rashba spin-orbit coupling (SOC) with a position-dependent orientation $\theta_{\mathrm{so}}(y,z)$ for which the normal direction is perpendicular to the $x$-axis.  We denote by $\mathbf{I}$ the identity matrix in the spin space and by $\sigma_{x,y,z}$ the Pauli matrices. Note that for the planar case we can simply set $\theta_{\mathrm{so}}(y,z) = 0$ and obtain the well-known two-dimensional form of SOC $\alpha\left(\sigma_x k_y - \sigma_y k_x\right)$. The third term of Eq.~(\ref{eq:Hsys}) describes the Zeeman coupling to an external magnetic field oriented along the $x$-axis given by $\mathbf{B}=(B,0,0)$, and having the vector potential $\mathbf{A}=(0,-zB,0)$. Here, $g$ is the Land\'e $g$-factor and $\mu_B$ the Bohr magneton. 

The numerical calculation of the conductance are performed with the KWANT software package \cite{Groth2014} using a nanowire of hexagonal shape and dimensions comparable to the experimental samples, following similar methodology as in the case of a recent publication \cite{Basaric2025}. The disorder potential is described by a Gaussian random field and fixed strength. The Hamiltonian (\ref{eq:Hsys}) is discretized and the magnetic field is implemented using the Peierls substitution. The nanowire is assumed to be in contact with two disorderless leads. 

This model is sufficient for characterizing the AB and AAS oscillations, as well as the phase rigidity present in higher-order harmonics. The magnetoconductance $G$ of the nanowire for fixed chemical potential $\mu$ and magnetic field was obtained using the Landauer-B\"uttiker formalism:
\begin{equation}
G(B,\mu)
=
\frac{e^{2}}{h}
\int dE\,
\left(-\frac{\partial \cal F}{\partial E}\right)
T(E,\mu,B) \ ,
\label{eq:LB}
\end{equation}
where $\cal F$ denotes the Fermi distribution, $E$ the energy, and $T$ the transmission function. For the temperature $T$ we assumed 1.6\,K. Additional parameters and specifications can be found in the following table:

\begin{table}[h]
\centering
\caption{Simulation parameters and specifications.}
\label{tab:sim_params}
\begin{tabular}{l c c}
\hline\hline
Parameter & Symbol & Value \\ 
\hline\hline
Effective mass & $m_{\mathrm{eff}}$ & 0.023\,$m_{\mathrm{e}}$  \\
Outer radius & $r_{\mathrm{out}}$ & 85\,nm  \\
Inner (core) radius & $r_{\mathrm{core}}$ & 55\,nm  \\
System length & $L$ & 0.6/2.0/4.0\,{\textmu}m \\
Rashba SOC parameter & $\alpha$ & 20 meV$\cdot$nm \\
Land\'e $g$-factor & $g$ & -14.9 \\
Chemical potential & $\mu$ & 10--50\,meV \\
Magnetic field & $B$ & -2 -- 2\,T \\
Lattice spacing & $a$ & 10\,nm \\
Temperature & $T$ & 1.6\,K \\
Correlation length & $l_{\mathrm{corr}}$ & 50\,nm \\
Disorder strength & $S_\mathrm{dis}$ & 8\,meV \\
\hline\hline
\end{tabular}
\end{table}

\section{Simulations with and without spin-orbit coupling}

\begin{figure*}[!h]
\centering
\vspace{0 mm} 
\includegraphics[width=0.6\linewidth]{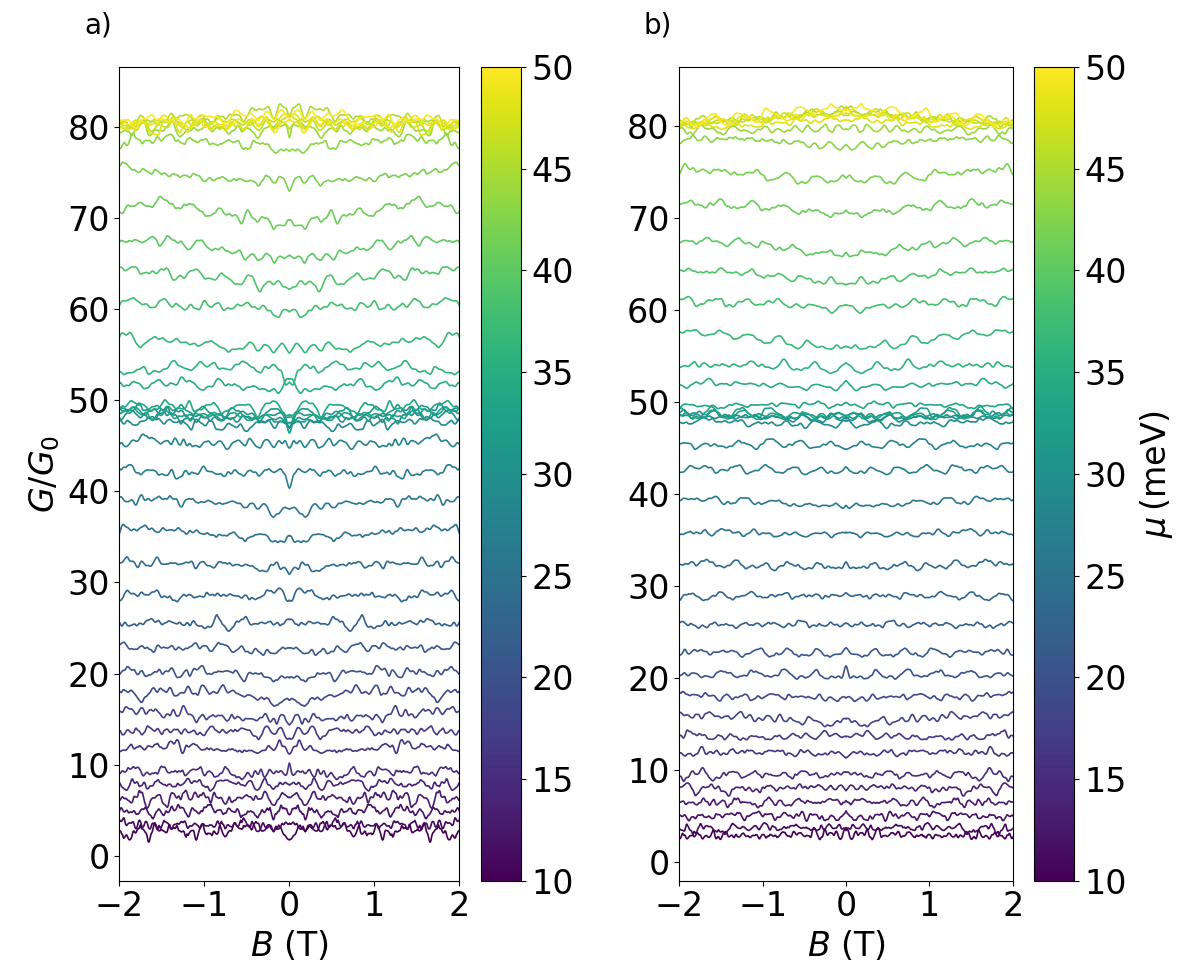}
\caption{\justifying Normalized conductance $G/G_0$ as function of magnetic field for chemical potentials $\mu$ ranging from 10 to 50\,meV for: (a) no spin-orbit coupling with $\alpha=0$, and (b) with included spin-orbit coupling with $\alpha=20$ meV$\cdot \,$nm.}
\label{fig:Cond}
\end{figure*}

In this section we show the results for the conductance computed using Eqs.~(\ref{eq:Hsys}) and (\ref{eq:LB}) for the nanowire explained above, together with its Fourier harmonics obtained with the FFT method. We begin by examining in Fig.~\ref{fig:Cond} the conductance of the nanowire of $L=2$ {\textmu}m for several values of the chemical potential, as a function of the magnetic field. The conductance curves show a dominant period of $h/e$-periodic magnetic flux, together with multiple irregularities, both without and with the spin-orbit interaction included. As discussed in the main text, there are several reasons for the irregularities, such as disorder or the internal geometry of the nanowire. A careful inspection around zero magnetic field indicates that for several chemical potentials the conductance curves have some systematic minima in the absence of SOC and some systematic maxima in the presence of it.

\begin{figure*}[!h]
\centering
\includegraphics[width=0.6\linewidth]{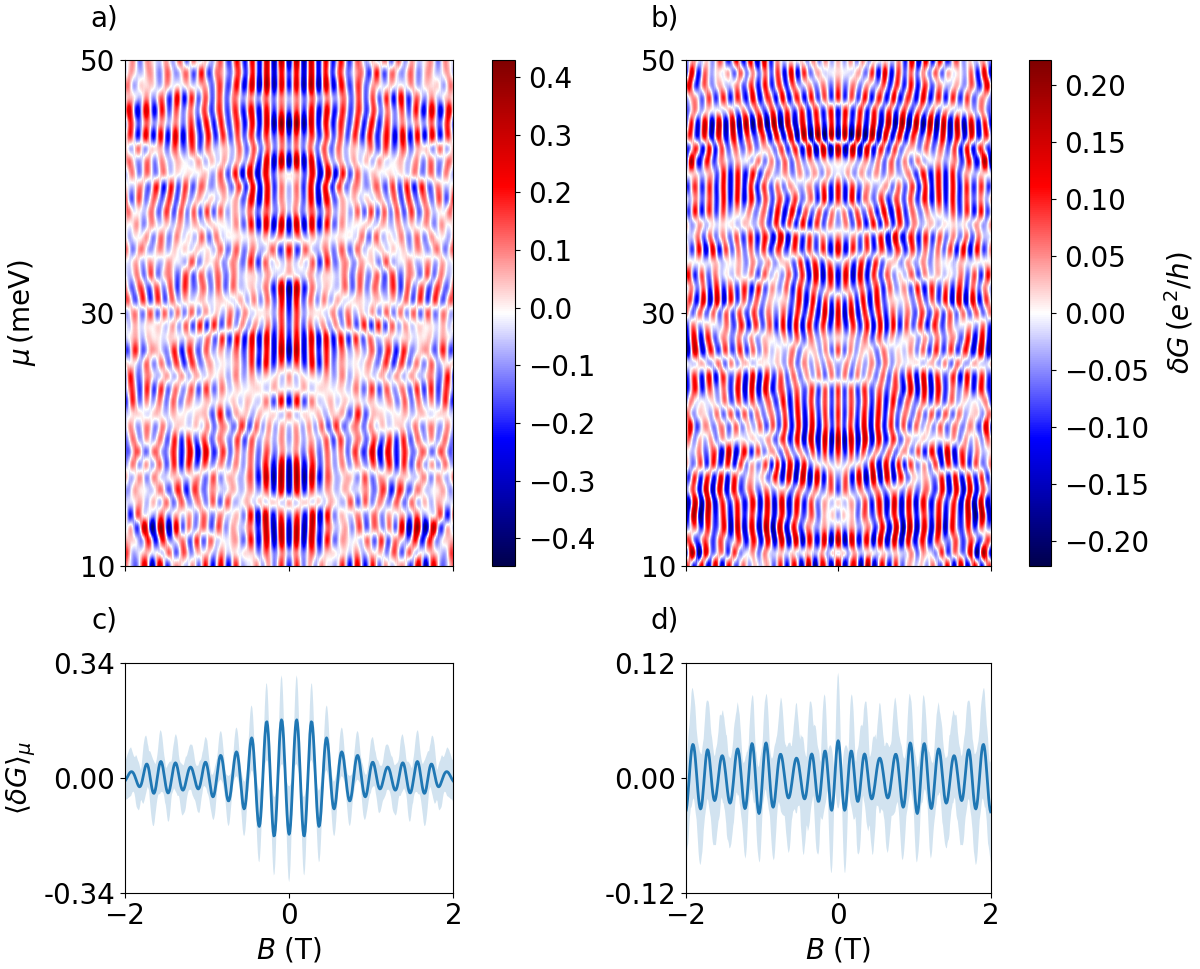}
\caption{\justifying Destructive (a), and constructive (b) wave interference for a wire with a length of 2\,{\textmu}m. The same disorder potential was used in both cases to ensure the existence of the AAS effect; however, in the case shown in (a), the spin-orbit constant $\alpha$ was set to 0, as opposed to 20\,meV$\cdot$nm in the case shown in (b). As can be seen from the averaged conductance in (c) and (d), for cases without and with spin-orbit coupling, the conductance changes from a minimum to a maximum at $B=0$. The respective standard deviations are also shown.}
\label{fig:aas}
\end{figure*}

In order to isolate the AAS effect, we examine the second harmonic of the oscillations, with a period of $h/2e$, shown in Fig.~\ref{fig:aas}. In the absence of SOC, we find that the conductance exhibits minima at zero magnetic field over large intervals of chemical potentials, as indicated by the blue vertical segments in Fig.\ref{fig:aas} (a). These segments correspond to the destructive interference of wave functions with their counterparts that are back-scattered by impurities. This phenomenon was originally predicted in disordered conductors without SOC \cite{Altshuler1981} and observed in ring arrays made of a material without spin-orbit coupling \cite{Dolan1986}. The phase rigidity with respect to the chemical potential continues over several periods of the magnetic flux. This can be seen in the color map as alternating blue and red lines and in the conductance averaged over all chemical potential values as a function of magnetic field, as shown in Fig.~\ref{fig:aas} (c). Thus, we obtain phase-rigidity of the magnetoconductance oscillations of destructive type in the absence of SOC, where the random potential included in the Hamiltonian (\ref{eq:Hsys}) plays a key role. 
  
Next, in the presence of SOC, Fig.~\ref{fig:aas} (b) shows how the minima of the conductance at zero field become maxima, i.e. the former blue and red lines switch colors. This observance is also supported by having a look at the averaged conductance shown in Fig.~\ref{fig:aas} (d), where a maximum is found at $B=0$. We interpret this result as the constructive interference of the wave functions mediated by the spin leading to phase-rigidity of the magnetoconductance observed in the experimental data shown in the  main text.  This is a similar relationship to that between the well-known weak localization and weak antilocalization effects in two-dimensional conductors \cite{Sharvin1981,Aronov1987}. Nevertheless, in our case the presence of SOC reduces the magnitude of the effect. Two important factors play a role in the phase rigidity of the $h/2e$ component. First one is the existence of the disorder potential, required for the existence of the AAS effect, and second is the spin-orbit strength represented by the $\alpha$ constant. The spin-orbit interaction of a sufficient strength is important because it gives rise to the constructive wave interference which in turn gives us the maximum of conductance at zero magnetic field.

\newpage

\section{Propagation of the phase rigidity to other harmonics}

In this section we will outline how the phase rigidity of the $h/2e$-periodic oscillations is propagated to the neighboring harmonics. Let us first write the conductance of a tubular nanowire as a discrete sum of Fourier components with period $\Phi_0$,
\begin{equation}
G(B) = g_0 + g_1 \cos\phi + g_2 \cos 2\phi + g_3\cos 3\phi + \ldots \ , 
\end{equation}
where $\phi=2\pi \Phi/\Phi_0$, $\Phi$ being the magnetic flux through the nanowire and $\Phi_0=h/e$ the flux quantum. Contributions due to interference of time-reversed paths such as the AAS effect are neglected so far. Because the conductance is an even function of the magnetic field $B$, i.~e. $G(B)=G(-B)$, the expansion contains only cosine functions. Since the conductance is not strictly periodic, for several obvious reasons such as the disorder inside the nanowire, the hexagonal geometry, the Zeeman spin splitting, etc., the Fourier amplitudes $g_n$ depend, more or less, on the magnetic field $B$.  However, let us assume that for low magnetic fields these coefficients have constant values. If so, for any $n > 1$:
\begin{equation}
g_n = \frac{1}{\pi} \int_{-\pi}^{\pi} G(B) \cos n\phi  \ d\phi \ .    
\label{eq:Fcoef}
\end{equation}
The presence of the AAS effect implies a specific contribution to the conductance of the form: 
\begin{equation}
G_{\textrm{AAS}}(B)=\gamma(B)\cos 2\phi \ , 
\label{eq:GAAS}
\end{equation}
and by adding such a correction the conductance becomes 
\begin{equation}
G^\prime (B) = G(B) + G_{\textrm{AAS}}(B) = G(B) + \gamma(B)\cos 2\phi = g^\prime_0 + g^\prime_1 \cos\phi + g^\prime_2 \cos 2\phi + g^\prime_3\cos 3\phi + \ldots \ ,   
\label{eq:Gprime}
\end{equation}
with the new Fourier coefficients being
\begin{equation}
g^\prime_n = \frac{1}{\pi} \int_{-\pi}^{\pi} G^\prime(B) \cos n\phi  \ d\phi = g_n + \frac{1}{\pi} \int_{-\pi}^{\pi} \gamma(B) \cos 2\phi \cos n\phi  \ d\phi \ .    
\end{equation}
Let us now admit that the AAS effect has a damping with the magnetic field, i.~e. it is only a low field phenomenon, as observed in the experimental data obtained here as well as in [\onlinecite{Basaric2025}], and qualitatively reproduced by the quantum mechanical calculations. If so, the amplitude $\gamma(B)$ is a decreasing function of $B$. The constructive interference of the wave functions implies $\gamma > 0$. A detailed analysis of this damping is beyond the scope of the present work. Instead, here we assume the simple empirical model:
\begin{equation}
\gamma(B) = \gamma_0 e^{-|\phi|/\beta} \ ,  
\label{eq:gamma}
\end{equation}
where $\gamma_0 = \gamma(0)$ and $\beta$ are positive constants.  With this model one obtains:
\begin{equation}
g^\prime_1 = g_1 + \frac{\gamma_0}{4\pi} \beta\left( e^{-\pi/\beta}+1 \right) 
\left( \frac{1}{\beta^2+1} + \frac{1}{9\beta^2+1} \right) \ ,      
\label{eq:Fcoeff1}
\end{equation}
and
\begin{equation}
g^\prime_3 = g_3 + \frac{\gamma_0}{4\pi} \beta\left( e^{-\pi/\beta}+1 \right) 
\left( \frac{1}{\beta^2+1} + \frac{1}{25\beta^2+1} \right) \ .     
\label{eq:Fcoeff3}
\end{equation}
We thus find positive corrections to the first and third harmonics of the conductance due to the fact the AAS contribution is an \textit{anharmonic} function of magnetic field. The relative effect on the first harmonic may be small, because of the large AB oscillations, but it may become large on the third harmonic which is obviously weaker.

\begin{figure*}[h]
\centering
\includegraphics[width=0.6\linewidth]{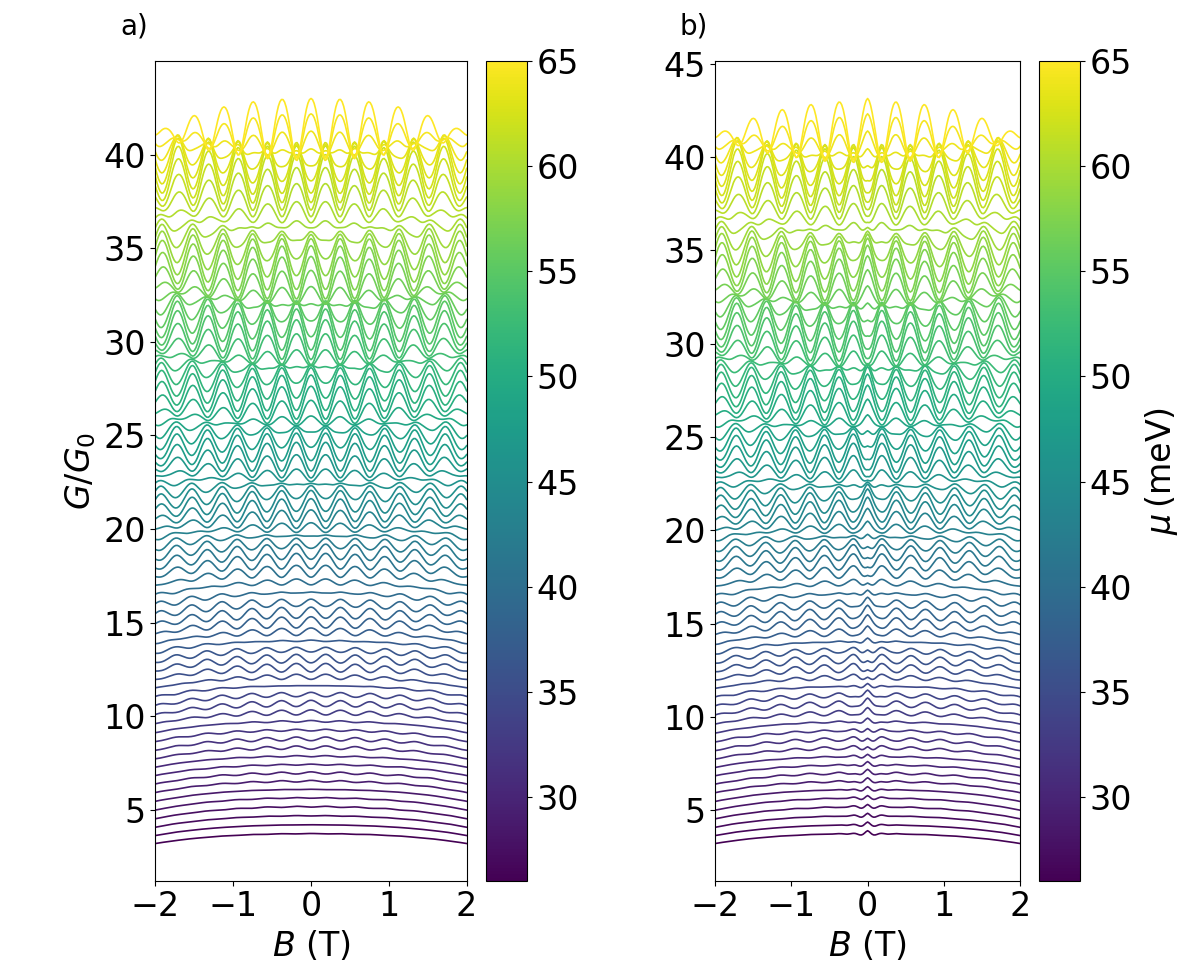}
\caption{\justifying (a) Conductance obtained with the Kubo formalism, where the AAS effect is not present. (b) Conductance with the AAS correction described by Eqs.~(\ref{eq:GAAS}), (\ref{eq:Gprime}), and (\ref{eq:gamma}) with parameters $\gamma_0=0.15$ and $\beta=2$.  }
\label{fig:Kubo}
\end{figure*}

To better demonstrate the propagation of the phase rigidity associated with the AAS effect into the other Fourier harmonics of the conductance, and especially in the third harmonic, we conducted the following numerical experiment. We used the conductivity of an infinite nanowire calculated with the Kubo formalism, where the scattering of electrons is described with the self-consistent Born approximation and $\delta$-form impurity potentials. This method does not capture effects related to weak localization such as AAS, but only the basic AB oscillations \cite{Basaric2025}. Still, due to the geometry and spin, higher harmonics are present and also the periodicity of the oscillations is deteriorated \cite{Ballester2013, Rosdahl2014, Urbaneja2018}. In Fig.~\ref{fig:Kubo} (a) we show the results for a GaAs/InAs core/shell hexagonal nanowire, with an external radius 79\,nm, a shell thickness 28\,nm, and infinite length. Next, in Fig.~\ref{fig:Kubo} (b) we show the results after adding to each conductivity value the AAS correction proposed in Eqs.~(\ref{eq:GAAS}) and (\ref{eq:gamma}), with $\gamma_0=0.15$ and $\beta=2$. The differences between the corrected and uncorrected results are small, with a small peak evolving around $B=0$ in the AAS-corrected Fig.~\ref{fig:Kubo} (b). 

\begin{figure*}[h]
\centering
\includegraphics[width=1.0\textwidth]{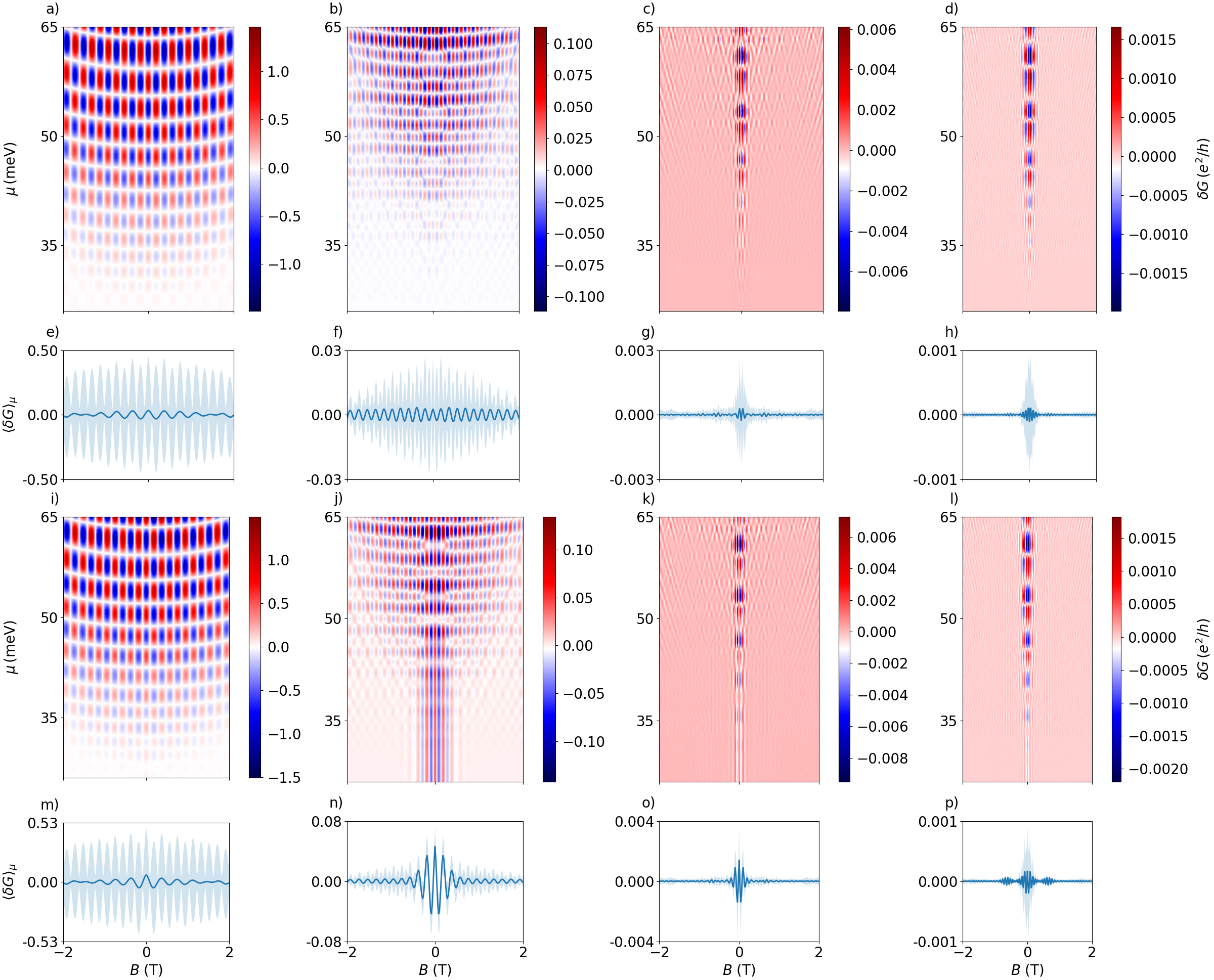}
\hfill
\caption{\justifying (a) - (d) The first four harmonics of the conductance obtained with the Kubo formalism with parameters $\gamma_0=0.15$ and $\beta=2$, where the AAS effect is not present. (e) - (h) Corresponding averaged conductance and standard deviation. (i) - (l) The first four harmonics of the conductance obtained with the Kubo formalism, with AAS effect present. (m) - (p) Corresponding averaged conductance and standard deviation.}
\label{fig:KF2}
\end{figure*}

However, if we look at the Fourier components shown in Fig.~\ref{fig:KF2}, the propagation of the phase rigidity becomes clear. The first harmonic of the uncorrected results has a positive mean value at zero magnetic field, which can be associated with the increase of the energy dispersion with increasing the chemical potential. However, we notice an increase of this mean value after adding the AAS correction, as predicted by Eq.~(\ref{eq:Fcoeff1}). In the second harmonic we see the obvious phase rigidity imposed by the $\cos(2\phi)$ factor of this correction. In the third harmonic we observe the phase rigidity propagation predicted by Eq.~(\ref{eq:Fcoeff3}).  Finally, we see it in the fourth harmonic as well. The phase rigidity trend of the fourth harmonic is not a surprise, since in reality the AAS correction should also have its own higher harmonics, i.~e. a term proportional to $\cos(4\phi)$, which we did not use in our model expressed by Eq.~(\ref{eq:GAAS}), but should be present in the experimental data. Indeed, by adding such a term to our model the phase rigidity of the fourth harmonic would become more evident.

\newpage

\section{Fast Fourier transforms of the \MakeLowercase{600\,nm} and \MakeLowercase{2\,{\textmu}m} long nanowires}

\begin{figure*}[h!]
\centering
\includegraphics[width=0.98\linewidth]{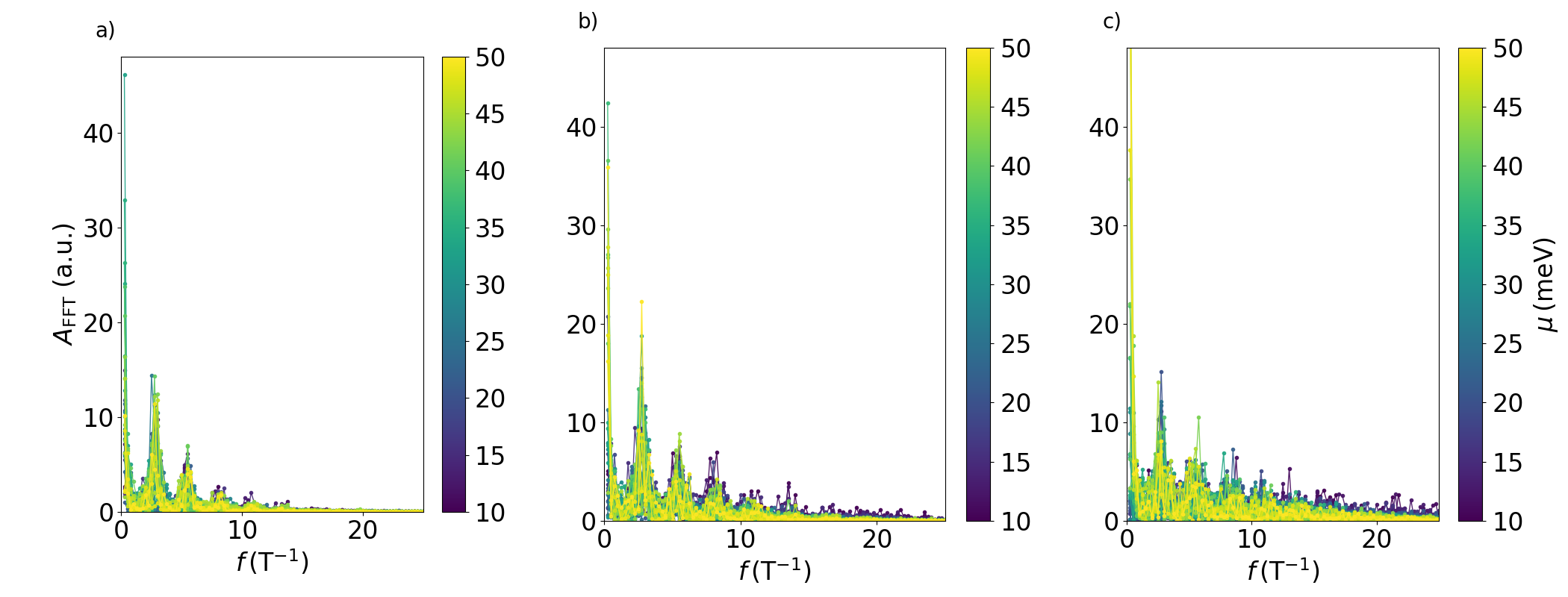}
\caption{\justifying  FFT spectra for the nanowires of length $L=600$ nm (a), $L=2$ {\textmu}m (b) and $L=4$ {\textmu}m (c) showing a more pronounced magnitude in the higher harmonics' peaks for the longer wire.}
\label{fig:FFT-sim-all}
\end{figure*}

Figures~\ref{fig:FFT-sim-all} (a) - (c) show the fast Fourier transforms of all calculated chemical potential values for nanowires that are 600\,nm, 2\,{\textmu}m and 4\,{\textmu}m long, respectively. The spectra are taken from the corresponding simulations shown in the main text. In all cases, the higher harmonics of the $h/e$-periodic oscillations are well resolved up to the fourth harmonics. The amplitude of the higher harmonics is larger for the longer nanowires compared to the shorter one.

\newpage

\putbib[bu2.bbl] 
\end{bibunit}

\end{document}